\documentclass[jsco]{academic} % NEW
\newenvironment{example}{\vskip 3pt \noindent {\bf Example:} }{}
\newcommand{\myvec}[1]{{\bf #1}}
\newenvironment{system}
{\refstepcounter{equation} %
\bgroup%
  $$%
  \openup\jot%
  \let\\=\mycr%
  \vcenter\bgroup%
  \tabskip=0pt plus 1fil%
\halign to \displaywidth
\bgroup\tabskip=0pt%
  $\hfil\displaystyle##$&%
  $\hfil{}##{}\hfil$&%
  $\displaystyle##$\hfil\tabskip=0.5in plus 1fil\cr%
  \noalign{\vskip-\jot}}%
{\cr\egroup\egroup%
\hspace{-0.4in}\hspace{-13pt}
\hspace{0.6in}
\llap{(\theequation)}$$\egroup\ignorespacesafterend} % 
\def\mycr{%
   {\ifnum0=`}\fi%
   \ifnum0=`{\fi}%
   \cr}%
\def\sech{\mathop{\mathrm{sech}}}
\def\tanh{\mathop{\mathrm{tanh}}}
\def\coth{\mathop{\mathrm{coth}}}
\def\sinh{\mathop{\mathrm{sinh}}}
\def\cosh{\mathop{\mathrm{cosh}}}
\def\cn{\mathop{\mathrm{cn}}}
\def\sn{\mathop{\mathrm{sn}}}
\def\dn{\mathop{\mathrm{dn}}}

\newcommand{\vecx}{\mbox{\bf x}}
\newcommand{\vecu}{\mbox{\bf u}}
\newcommand{\vecU}{\mbox{\bf U}}
\begin{document} % NEW
\shortauthor{D.\ Baldwin {\em et al.}}
\shorttitle{Symbolic computation of Tanh, Sech, Cn, and Sn Solutions}
\title{{
Symbolic computation of exact solutions expressible in hyperbolic and 
elliptic functions for nonlinear PDEs}
\footnote{This material is based upon research supported by the 
National Science Foundation under Grants 
Nos.\ DMS-9732069, DMS-9912293 and CCR-9901929. \\
$\phantom{.}\;\;\;\star$Correspondence: whereman@mines.edu } }
\author[1]{D.\ Baldwin}
\author[2]{\"{U}.\ G\"{o}kta\c{s}}
\author[1,3]{W.\ Hereman}
\author[4]{L.\ Hong}
\author[1]{R.\ S.\ Martino}
\author[5]{J.\ C.\ Miller}
\address[1]{Department of Mathematical and Computer Sciences,
Colorado School of Mines, Golden, CO 80401-1887, U.S.A. }
\address[2]{Wolfram Research, Inc., 100 Trade Center Drive, 
Champaign, IL 61820, U.S.A.}
\address[3]{Department of Applied Mathematics, 
University of Stellenbosch, Private Bag X1, 7602 Matieland, South Africa}
\address[4]{Department of Engineering and Applied Sciences, 
Harvard University, Cambridge, MA 02138, U.S.A.}
\address[5]{Department of Applied Mathematics and Theoretical Physics,
Churchill College, University of Cambridge, Cambridge CB3 0DS, England, U.K.}
\label{firstpage}
\maketitle
\vskip 1pt
\noindent
% 
% ABSTRACT
% 
\begin{abstract}
Algorithms are presented for the tanh- and sech-methods, which lead to 
closed-form solutions of nonlinear ordinary and partial differential 
equations (ODEs and PDEs). 
New algorithms are given to find exact polynomial solutions of ODEs 
and PDEs in terms of Jacobi's elliptic functions. 

For systems with parameters, the algorithms determine the 
conditions on the parameters so that the differential equations 
admit polynomial solutions in tanh, sech, combinations thereof,  
Jacobi's sn or cn functions.
Examples illustrate key steps of the algorithms. 

The new algorithms are implemented in {\em Mathematica}.
The package {\rm PDESpecialSolutions.m} can be used to automatically compute 
new special solutions of nonlinear PDEs. 
Use of the package, implementation issues, scope, limitations, and 
future extensions of the software are addressed. 

A survey is given of related algorithms and symbolic software to
compute exact solutions of nonlinear differential equations.
\end{abstract}
% 
%%%%%%%%%%%%%%%%%%%%%%%%%%%Introduction%%%%%%%%%%%%%%%%%%%%%%%%%%%%
% INTRODUCTION
% 
% \vspace{-1.25cm}
\section{Introduction}
\label{intro}
The appearance of solitary wave solutions in nature is quite common.
Bell shaped sech-solutions and kink shaped tanh-solutions model wave 
phenomena in fluids, plasmas, elastic media, 
electrical circuits, optical fibers, chemical reactions, bio-genetics, etc.
The travelling wave solutions of the Korteweg-de Vries (KdV) and Boussinesq 
equations, which describe water waves, are famous examples. 

Apart from their physical relevance, the knowledge of closed-form solutions 
of nonlinear ordinary and partial differential equations (ODEs and PDEs) 
facilitates the testing of numerical solvers, and aids in the stability 
analysis.
Indeed, the exact solutions given in this paper correspond to homoclinic 
and heteroclinic orbits in phase space, which are the separatrices of 
stable and unstable regions. 

Travelling wave solutions of many nonlinear ODEs and PDEs from 
soliton theory (and beyond) can often be expressed as polynomials of the 
hyperbolic tangent and secant functions. 
An explanation is given in, for example, Hereman and Takaoka (1990). 
The existence of solitary wave solutions of evolution equations is 
addressed in Kichenassamy and Olver (1993).
The tanh-method provides a straightforward algorithm to compute such 
particular solutions for a large class of nonlinear PDEs. 
Consult Malfliet (1992, 2003), Malfliet and Hereman (1996), 
and Das and Sarma (1999) for a multitude of references to tanh-based 
techniques and applications. 

The tanh-method for, say, a single PDE in $u(x,t)$ works as follows: 
In a travelling frame of reference, $\xi = c_1 x + c_2 t + \Delta,$ 
one transforms the PDE into an ODE in the new independent variable 
$T = {\tanh}\,\xi.$ 
Since the derivative of $\tanh$ is {\rm polynomial} in $\tanh,$ 
i.e., $T^{\prime}= 1-T^2,$ all derivatives of $T$ are polynomials of $T.$
Via a chain rule, the polynomial PDE in $u(x,t)$ is transformed into 
an ODE in $U(T),$ which has polynomial coefficients in $T$. 
One then seeks polynomial solutions of the ODE, thus generating a subset 
of the set of all solutions. 

Along the path, one encounters ODEs which are nonlinear, higher-order 
versions of the ultraspherical differential equation, 
\begin{equation}
\label{ultraspherical}
(1 - x^2) y^{\prime\prime}(x) - 
(2 \alpha + 1) x y^{\prime}(x) + n (n + 2 \alpha) y(x) = 0,
\end{equation}
with integer $n \geq 0$ and $\alpha$ real, whose solutions are the 
Gegenbauer polynomials. 
Eq.\ (\ref{ultraspherical}) includes the Legendre equation 
$(\alpha = \textstyle{\frac{1}{2}}),$ satisfied by the Legendre polynomials, 
and the ODEs for Chebeyshev polynomials of type I $(\alpha = 0)$ 
and type II $(\alpha = 1).$ 
Likewise, the associated Legendre equation, 
\begin{equation}
\label{associatedlegendre}
(1 - x^2)^2 y^{\prime\prime}(x) - 
2 x (x^2 -1) y^{\prime}(x) + [n(n+1)(1-x^2)- m^2] y(x) = 0,
\end{equation}
with $m$ and $n$ non-negative integers, appears in solving the 
Sturm-Liouville problem for the KdV with a sech-square potential 
(see Drazin and Johnson, 1989). 

The appeal and success of the tanh-method lies in the fact that one 
circumvents integration to get explicit solutions. 
Variants of the method appear in mathematical physics,
plasma physics, and fluid dynamics.  
For early references see e.g.\ Malfliet, 1992; Yang, 1994; and
Das and Sarma, 1999. 
Recently, the tanh-methods have been applied to many nonlinear PDEs 
in multiple independent variables
(see Fan, 2002abc, 2003abc; Fan and Hon, 2002, 2003ab; Gao and Tian, 2001; 
Li and Liu, 2002; Yao and Li, 2002ab). 

In this paper we present three flavors of tanh- and sech-methods as they 
apply to nonlinear polynomial systems of ODEs and PDEs.
Based on the strategy of the tanh-method, we also present algorithms to 
compute polynomial solutions in terms of the Jacobi $\sn$ and 
$\cn$ functions. 
Applied to the KdV equation, the so-called cnoidal solution 
(Drazin and Johnson, 1989) is obtained. 
For Duffing's equation (Lawden, 1989), we recover known $\sn$ 
and cn-solutions which model vibrations of a nonlinear spring. 
Sn- and cn-methods are quite effective for symbolically solving 
nonlinear PDEs as shown in Fu {\em et al.\/} (2001), 
Parkes {\em et al.\/} (2002), Liu and Li (2002ab), Fan and Zhang (2002), 
Fan (2003abc), Chen and Zhang (2003ab), and Yan (2003).

We also present our package, 
\verb|PDESpecialSolutions.m| (Baldwin {\em et al.}, 2001) 
in {\em Mathematica}, which implements the five methods.  
Without intervention by the user, our software computes travelling wave 
solutions as polynomials in either $T = {\tanh}\,\xi$, $S={\sech}\,\xi,$ 
combinations thereof, $\mbox{CN} = \cn(\xi;m)$, or $\mbox{SN} = \sn(\xi;m)$ 
with $\xi = c_1 x + c_2 y + c_3 z + \dots + c_n t + \Delta = 
\sum_{j=0}^N c_j x_j + \Delta.$
The coefficients of the spatial coordinates are the {\rm components of 
the wave vector}; the time coefficient is the {\rm angular frequency} 
of the wave. 
The wave travels in the direction of the wave vector; 
its plane wave front is perpendicular to that wave vector. 
$\Delta$ is the constant phase. 
For systems of ODEs or PDEs with constant parameters, the software 
automatically determines the conditions on the parameters so that the 
equations might admit polynomial solutions in 
$\tanh, \sech,$ both, $\sn$ or $\cn.$

Parkes and Duffy (1996) mention the difficulty of using the tanh-method 
by hand for anything but simple PDEs. 
Therefore, they automated to some degree the tanh-method using 
{\em Mathematica}. 
Their code {\rm ATFM} carries out some (but not all) steps of the method.
Parkes {\em et al.} (1998) also considered solutions to (odd-order 
generalized KdV) equations in even powers of $\sech.$ 
The code {\rm ATFM} does not cover solutions involving odd powers of $\sech.$
Recently, Parkes {\em et al.}\/ (2002) extended their methods to cover the 
Jacobi elliptic functions. 
Abbott {\it et al.\/} (2002) produced the function {\rm SeriesSn} to 
partially automate the elliptic function method.
Li and Liu (2002) designed the {\em Maple} package {\rm RATH} to automate 
the tanh-method. 
In Liu and Li (2002a) they announce their {\em Maple} code {\rm AJFM} for
the Jacobi elliptic function method. 
In Section~\ref{reviewsoftware} we review the codes {\rm ATFM}, 
{\rm RATH}, {\rm AJFM}, and {\rm SeriesSn} and compare them with 
{\rm PDESpecialSolutions.m}. 

The paper is organized as follows:
In Sections~\ref{tanhmethodPDEs} and~\ref{sechmethodPDEs}, we give the 
main steps of the algorithms for computing tanh- and sech-solutions of 
nonlinear polynomial PDEs. 
We restrict ourselves to polynomial solutions in either $\tanh$ or $\sech.$
The Boussinesq equation and Hirota-Satsuma system of coupled KdV 
equations illustrate the steps. 
For references to both equations see e.g.\ Ablowitz and Clarkson (1991).
In Section~\ref{sechtanhmethodPDEs} we consider a broader class of
polynomial solutions involving both $\tanh$ and $\sech.$ 
The tanh-sech algorithm is used to solve a system of PDEs due 
to Gao and Tian (2001). 
In Section~\ref{cnmethodPDEs} we show how modifying the chain rule 
allows us to find polynomial solutions in cn and sn.
The KdV equation is used to illustrates the steps.
In Section~\ref{algorithms} we give details of the algorithms to compute 
the highest-degree of the polynomials, to analyze and solve 
nonlinear algebraic systems with parameters, and to numerically 
and symbolically test solutions.
The coupled KdV equations illustrate the subtleties of these algorithms.
In Section~\ref{examplesPDEs} we present exact solutions for several
nonlinear ODEs and PDEs. 
In Section~\ref{relatedalgossoftware} we address other perspectives and 
extensions of the algorithms, and review related software packages.
We discuss the results and draw some conclusions in Section~\ref{conclusions}.
The use of the package  {\rm PDESpecialSolutions.m} is shown in Appendix A.
% 
%%%%%%%%%%%%%%%%%%Tanh method for nonlinear PDEs%%%%%%%%%%%%%%%%%%%%%%%%%%%%
% 
\section{Algorithm to compute tanh-solutions for nonlinear PDEs}
\label{tanhmethodPDEs}

In this section we outline the tanh-method (Malfliet and Hereman, 1996) 
for the computation of closed-form $\tanh$-solutions for nonlinear 
PDEs (and ODEs). 
Each of the five main steps of our algorithm is illustrated for the 
Boussinesq equation. 
Details of steps T2, T4 and T5 are postponed to Section~\ref{algorithms}.

Given is a system of polynomial PDEs with constant coefficients, 
\begin{equation}
\label{originalsystemPDEs}
{\bf \Delta} ( {\vecu}({\vecx}), {\vecu}^{\prime}({\vecx}), 
{\vecu}^{\prime\prime}({\vecx}), \cdots, {\vecu}^{(k)} ({\vecx}),
\cdots, {\vecu}^{(m)} ({\vecx}) ) = {\bf 0}, 
\end{equation}
where the dependent variable ${\vecu}$ has $M$ components $u_i,$ 
the independent variable ${\vecx}$ has $N$ components $x_j,$ 
and ${\vecu}^{(k)}({\vecx})$ denotes the collection of mixed 
derivative terms of order $k.$ 
Lower-case Greek letters will denote parameters in (\ref{originalsystemPDEs}).

For notational simplicity, in Section~\ref{examplesPDEs} we will use 
dependent variables $u, v, w,$ etc.\ and independent variables 
$x, y, z,$ and $t.$
\vskip 2pt
\noindent
{\bf Example:} The classical Boussinesq equation,
\begin{equation}
\label{orgboussinesq}
u_{tt} - u_{xx} + 3 u u_{xx} + 3 u_x^2 + \alpha u_{xxxx} = 0,
\end{equation}
with real parameter $\alpha,$ was proposed by Boussinesq to describe 
surface water waves whose horizontal scale is much larger than the depth 
of the water (Ablowitz and Clarkson, 1991). 
Variants of (\ref{orgboussinesq}) were recently solved by Fan and Hon (2003a). 

While one could apply the tanh-method directly to (\ref{orgboussinesq}),  
we recast it as a first order system in time to show the method for a 
simple system of PDEs. So, 
\begin{system}
\label{boussinesq}
&& u_{1,x_2} + u_{2,x_1} = 0, \\ 
&& u_{2,x_2} + u_{1,x_1} - 3 u_1 u_{1,x_1} - \alpha u_{1,3 x_1} = 0,
\end{system}
where $x_1 = x, x_2 = t, u_1(x_1,x_2)=u(x,t),$ and $u_2(x_1,x_2)= u_t(x,t).$
We use
\begin{equation}
\label{derivativenotation} 
u_{i, k x_j} \buildrel {\rm def} \over = 
\frac{\partial^k {u_i}}{\partial{x_j^k}}, \quad
u_{i, p x_j \, r x_k \, s x_\ell} \buildrel {\rm def} \over = 
\frac{\partial^{p+r+s} {u_i}}
{\partial{x_j^p}\partial{x_k^r}\partial{x_\ell^s}}, \; {\rm etc.}
\end{equation}
through out this paper.
\vfill
\newpage
\noindent
{\bf Step T1:$\;\;\;$Transform the PDE into a nonlinear ODE}
\vskip 2pt
\noindent
We seek solutions in the travelling frame of reference, 
\begin{equation}
\label{framePDE}
\xi = \sum_{j=1}^{N} c_j x_j + \Delta, 
\end{equation}
where $c_j$ and $\Delta$ are constant.

The tanh-method seeks polynomial solutions expressible in the 
hyperbolic tangent, $T = {\tanh}\,\xi.$ 
Based on the identity
$ {\cosh}^2\xi - {\sinh}^2\xi = 1 $
one computes
\begin{system}
\label{tanhderivative}
{\tanh}^{\prime}\xi &=& {\sech}^2\xi = 1-{\tanh}^2\xi, \\
{\tanh}^{\prime\prime}\xi &=& 
  -2 \, {\tanh}\,\xi + 2 \, {\tanh}^3\xi, \;\; {\rm etc.}
\end{system}
Therefore, the first and, consequently, all higher-order derivatives are 
polynomials in $T.$
Since $T^{\prime}=1-T^2,$ repeatedly applying the chain rule, 
\begin{equation}
\label{chainruletanhPDE}
\frac{\partial \bullet}{\partial x_j} 
= \frac{d\bullet}{dT} \frac{dT}{d\xi} \frac{\partial \xi}{\partial x_j} 
= c_j (1-T^2) \frac{d \bullet}{dT}, 
\end{equation}
transforms the system of PDEs into a coupled system of nonlinear ODEs, 
\begin{equation}
\label{legendretypetanhPDEs}
{\bf \Delta}(T, {\vecU}(T), {\vecU}^{\prime}(T), {\vecU}^{\prime\prime}(T), 
\ldots, {\vecU}^{(m)}(T)) = {\bf 0}, 
\end{equation}
with ${\vecU}(T) = {\vecu}({\vecx}).$ 
Each component of ${\bf \Delta}$ is a nonlinear ODE with polynomial 
coefficients in $T.$  
\vskip 2pt
\noindent
{\bf Example:} Substituting 
\begin{system}
\label{derivativesxitanhPDEs}
u_{i, x_j}  &=& c_j (1-T^2) U_i^{\prime}, \\
u_{i, 2x_j} &=& c_j^2 (1-T^2) \left[(1-T^2) U_i^{\prime} \right]^{\prime} 
= c_j^2 (1-T^2) [-2T U_i^{\prime} + (1-T^2) U_i^{\prime\prime}], \\
u_{i, 3x_j} &=& 
 c_j^3 (1-T^2) \left[-2T (1-T^2)U_i^{\prime}
 + (1-T^2)^2 U_i^{\prime\prime} \right]^{\prime} \\
 &=& c_j^3 (1-T^2)
 \left[-2(1-3T^2)U_i^{\prime}-6T (1-T^2) U_i^{\prime\prime}
 + (1-T^2)^2 U_i^{\prime\prime\prime}\right], 
\end{system}
into (\ref{boussinesq}), and cancelling common $(1-T^2)$ factors, yields
\begin{system}
\label{boussinesqlegendre}
&&c_2 U_1^{\prime}+c_1 U_2^{\prime}=0, \\
&&c_2 U_2^{\prime}\!+\!c_1U_1^{\prime}\!-\!\!3c_1U_1U_1^{\prime}
\!+\!\alpha c_1^3
\left[2(1\!\!-\!3T^2)U_1^{\prime}\!+\!6T(1\!\!-\!T^2)U_1^{\prime\prime}\!-
\!\!(1\!\!-\!T^2)^2U_1^{\prime\prime\prime}\right]\!=\!0,
\end{system}
where $U_1(T)=u_1(x_1,x_2)$ and $U_2(T)=u_2(x_1,x_2).$
\vskip 2pt
\noindent
{\bf Step T2:$\;\;\;$Determine the degree of the polynomial solutions}
\vskip 2pt
\noindent
Seeking polynomial solutions of the form
\begin{equation}
\label{polynomialsolutiontanhPDEs}
U_i(T) = \sum_{j=0}^{M_i} a_{ij} T^j,
\end{equation}
\vskip 1pt
\noindent
we must determine the leading exponents $M_i$ before the $a_{ij}$ can 
be computed.
We assume that $M_i \geq 1$ to avoid trivial solutions.
Substituting $U_i$ into (\ref{legendretypetanhPDEs}), the coefficients of 
every power of $T$ in every equation must vanish.
In particular, the highest degree terms must vanish.
Since the highest degree terms depend only on $T^{M_i}$ in 
(\ref{polynomialsolutiontanhPDEs}), it suffices to substitute 
$U_i (T) = T^{M_i}$ into the left hand side of (\ref{legendretypetanhPDEs}). 
In the resulting polynomial system ${\bf P}(T),$
equating every two possible highest exponents in every component $P_i$
gives a linear system for the $M_i.$ 
That linear system is then solved.

If one or more exponents $M_i$ remain undetermined, assign an integer value
to the free $M_i$ so that every equation in (\ref{legendretypetanhPDEs}) has 
at least two different terms with equal highest exponents. 
Carry each solution to step T3.
\vskip 2pt
\noindent
{\bf Example:} For the Boussinesq system, substituting 
$U_1(T) = T^{M_1}$ and $U_2(T) = T^{M_2}$ into 
(\ref{boussinesqlegendre}), and equating the highest exponents of $T$ 
for each equation, gives
\begin{equation}
\label{boussinesqpowerbalance}
M_1 - 1 = M_2 - 1, \quad 2 M_1 - 1 = M_1 + 1.
\end{equation}
Then, $M_1 = M_2 = 2,$ and
\begin{equation}
\label{boussinesqpols}
U_1(T) = a_{10} + a_{11} T + a_{12} T^2,  \;\;
U_2(T) = a_{20} + a_{21} T + a_{22} T^2. 
\end{equation}
\vskip .005pt
\noindent
{\bf Step T3:$\;\;\;$Derive the algebraic system for the coefficients $a_{ij}$}
\vskip 2pt
\noindent 
To generate the system for the unknown coefficients $a_{ij}$ and wave 
parameters $c_j$, substitute (\ref{polynomialsolutiontanhPDEs}) 
into (\ref{legendretypetanhPDEs}) and set the coefficients of $T^i$ to zero.
The resulting nonlinear algebraic system for the unknowns $a_{ij}$ 
is parameterized by the $c_j,$ and the external parameters 
(in lower-case Greek letters) of system (\ref{originalsystemPDEs}), if any. 
\vskip 2pt
\noindent
{\bf Example:} 
Continuing with the Boussinesq system, after substituting 
(\ref{boussinesqpols}) into (\ref{boussinesqlegendre}), 
and collecting the terms of like degree in $T,$ 
we get (in order of complexity):
\begin{system}
\label{boussinesqalgsys}
a_{21} \, c_1 + a_{11}  \, c_2  &=& 0,  \\ 
a_{22} \, c_1 + a_{12} \, c_2  &=& 0,   \\ 
a_{11} \, c_1 \, (3 a_{12} + 2 \alpha \, c_1^2) &=& 0,  \\
a_{12} \, c_1 \, (a_{12} + 4 \alpha  \, c_1^2) &=& 0,   \\ 
a_{11}  \, c_1 - 3  a_{10} \, a_{11}  \, c_1 + 
2 \alpha  a_{11}  \, c_1^3 + a_{21} \, c_2 &=& 0,   \\ 
-3 a_{11}^2 \, c_1 + 2\, a_{12} \, c_1 -6 a_{10} \, a_{12} \, c_1 
+ 16 \alpha \, a_{12} \, c_1^3 + 2 a_{22} \, c_2  &=& 0, 
\end{system}
with unknowns $a_{10}, a_{11}, a_{12}, a_{20}, a_{21}, a_{22},$ and 
parameters $c_1,c_2,$ and $\alpha.$
\vskip 2pt
\noindent
{\bf Step T4:$\;\;\;$Solve the nonlinear parameterized algebraic system}
\vskip 2pt
\noindent
The most difficult step is solving the nonlinear algebraic system. 
To do so, we designed a customized, yet powerful, nonlinear solver
(see Section~\ref{analyzeandsolve} for details). 

The nonlinear algebraic system is solved with the following assumptions: 
\vskip 0.50pt
\noindent
(i) all parameters, $\alpha, \beta,$ etc., in (\ref{originalsystemPDEs}) 
are strictly positive. 
Vanishing parameters may change the exponents $M_i$ in step T2.
To compute solutions corresponding to negative parameters, reverse the 
signs of the parameters in the PDE. 
For example, replace $\alpha$ by $-\alpha$ in (\ref{orgboussinesq}).
\vskip 1pt
\noindent
(ii) the coefficients of the highest power terms
$(a_{i \, M_i}, \; i=1,\cdots,M)$ 
in (\ref{polynomialsolutiontanhPDEs}) are all nonzero 
(for consistency with step T2).
\vskip 1pt
\noindent
(iii) all $c_j$ are nonzero (demanded by the physical nature of the solutions).
\vskip 2pt
\noindent
{\bf Example:} 
Assuming $c_1, c_2, a_{12}, a_{22},$ and $\alpha$ nonzero, the solution 
of (\ref{boussinesqalgsys}) is
\begin{system}
\label{boussinesqalgsol}
a_{10} &=& (c_1^2 -c_2^2 + 8 \alpha c_1^4)/(3 c_1^2), \;
a_{11} = 0, \;\; 
a_{12} = - 4 \alpha c_1^2, \\
a_{20} &=& {\rm arbitrary}, \;\;
a_{21} = 0, \;\;
a_{22} = 4 \alpha c_1 c_2.
\end{system}
In this case, there are no conditions on the parameters $c_1, c_2$ and 
$\alpha.$ 
\vskip 2pt
\noindent
{\bf Step T5:$\;\;\;$Build and test the solitary wave solutions} 
\vskip 2pt
\noindent
Substitute the solutions obtained in step T4 into 
(\ref{polynomialsolutiontanhPDEs}) and reverse step T1
to obtain the explicit solutions in the original variables.
It is prudent to test the solutions by substituting them into 
(\ref{originalsystemPDEs}). 
For details about testing see Section~\ref{testing}. 
\vskip 2pt
\noindent
{\bf Example:} Inserting (\ref{boussinesqalgsol}) into 
(\ref{boussinesqpols}), and replacing 
$T = {\tanh}(c_1 x + c_2 t + \Delta),$  
the closed form solution for (\ref{boussinesq}) 
(or (\ref{orgboussinesq})) is 
\begin{system}
\label{boussinesqtanhsolution}
u(x,t) &=& u_1(x,t) = {(c_1^2 - c_2^2 + 8 \alpha c_1^4)}/{(3 c_1^2)} 
- 4 \alpha c_1^2 \, {\tanh}^2(c_1 x + c_2 t + \Delta),   \\ 
u_2(x,t) &=& -{\textstyle \int} u_{1,t}(x,t) dx = a_{20} + 4 \alpha c_1 c_2 
\, {\tanh}^2(c_1 x + c_2 t + \Delta), 
\end{system}
where $a_{20}, c_1, c_2, \alpha$ and $\Delta$ are arbitrary. 
Steps T1-T5 must be repeated if one or more of the external 
parameters (lower-case Greeks) are set to zero. 
% 
%%%%%%%%%%%%%%%%%%%Sech method for nonlinear PDEs%%%%%%%%%%%%%%%%%%%%%%
% 
\vspace{-0.25cm}
\section{Algorithm to compute sech-solutions for nonlinear PDEs}
\label{sechmethodPDEs}

In this section we restrict ourselves to polynomial solutions of 
(\ref{originalsystemPDEs}) in $\sech.$ 
Polynomial solutions involving both $\sech$ and $\tanh$ are dealt with in 
Section~\ref{sechtanhmethodPDEs}. 
Details of the algorithms for steps S2, S4 and S5 are in 
Section~\ref{algorithms}.

Using
$ {\tanh}^2\xi + {\sech}^2\xi = 1, $
solution (\ref{boussinesqtanhsolution}) of (\ref{boussinesq}) can be 
expressed as
\begin{system}
\label{boussinesqsechsolution}
u_1(x,t) &=& {(c_1^2 - c_2^2 - 4 \alpha c_1^4)}/{(3 c_1^2)} +
4 \alpha c_1^2 \, {\sech}^2(c_1 x + c_2 t + \Delta),  \\ 
u_2(x,t) &=& a_{20} + 4 \alpha c_1 c_2 -  
4 \alpha c_1 c_2 \, {\sech}^2(c_1 x + c_2 t + \Delta).
\end{system}
Obviously, any even order solution in $\tanh$ can be written in even orders 
of $\sech.$ 
Some PDEs however have polynomial solutions of odd-order in $\sech.$
For example, the modified KdV equation (Ablowitz and Clarkson, 1991),
\begin{equation}
\label{mkdv}
u_t + \alpha u^2 u_x + u_{xxx} = 0, 
\end{equation}
has the solution 
\begin{equation}
\label{mkdvsechsolution}
u(x,t) = \pm c_1 \sqrt{{6}/{\alpha}} 
         \, {\sech}(c_1 x - c_1^3 t + \Delta), 
\end{equation}
which cannot be found using the tanh-method. 
\vskip 2pt
\noindent
{\bf Example:} 
The five main steps of the $\sech$-algorithm are illustrated with 
the Hirota-Satsuma system of coupled KdV equations 
(Ablowitz and Clarkson, 1991),  
\begin{system}
\label{orgckdv}
&& u_t - \alpha (6 u u_x + u_{xxx}) + 2 \beta v v_x = 0,  \\
&& v_t + 3 u v_x + v_{xxx} = 0,
\end{system}
with real parameters $\alpha, \beta.$ 
Sech-type solutions were reported in Hereman (1991) and Fan and Hon (2002).
Variants and generalizations of (\ref{orgckdv}) were solved in 
Chen and Zhang (2003a) and Yan (2003).

Letting $u_1(x_1,x_2) = u(x,t)$ and $u_2(x_1,x_2) = v(x,t),$
Eq.\ (\ref{orgckdv}) is then 
\begin{system}
\label{ckdv}
&& u_{1,x_2} - \alpha (6 u_{1} u_{1,x_1} + u_{1,3x_1}) 
+ 2 \beta u_{2} u_{2,x_1} = 0, \\
&& u_{2,x_2} + 3 u_{1} u_{2,x_1} + u_{2,3x_1} = 0.
\end{system}
\vskip .005pt
\noindent
{\bf Step S1:$\;\;\;$Transform the PDE into a nonlinear ODE}
\vskip 2pt
\noindent
Adhering to the travelling frame of reference (\ref{framePDE}),  
and using ${\tanh}^2\xi + {\sech}^2\xi = 1,$
\begin{equation}
\label{sechderivative}
{\sech}^{\prime}\xi = - {\sech}\,\xi \, {\tanh}\,\xi 
= -{\sech}\,\xi \sqrt{1 - {\sech}^2\xi}. 
\end{equation}
Setting $S={\sech}\,\xi$ and repeatedly applying the chain rule, 
\begin{equation}
\label{chainrulesechPDE}
\frac{\partial \bullet}{\partial x_j} 
= \frac{d\bullet}{dS} \frac{dS}{d\xi} \frac{\partial \xi}{\partial x_j} 
= - c_j S \sqrt{1-S^2} \frac{d \bullet}{dS}, 
\end{equation}
(\ref{originalsystemPDEs}) is transformed into a system of nonlinear 
ODEs of the form: 
\begin{equation}
\label{generaltypesech}
{\bf \Gamma}(S, {\vecU}(S), {\vecU}^{\prime}(S), \ldots ) + \sqrt{1-S^2} 
\, {\bf \Pi}(S, {\vecU}(S), {\vecU}^{\prime}(S), \ldots ) = {\bf 0}, 
\end{equation}
where ${\vecU}(S) = {\vecu}({\vecx}),$ and all components of 
${\bf \Gamma}$ and ${\bf \Pi}$ are ODEs with polynomial coefficients in $S.$  
If either ${\bf \Gamma}$ or ${\bf \Pi}$ is identically ${\bf 0},$ then
\begin{equation}
\label{legendretypesech}
{\bf \Delta}(S, {\vecU}(S), {\vecU}^{\prime}(S), \ldots ) = {\bf 0},
\end{equation}
where ${\bf \Delta}$ is either ${\bf \Gamma}$ or ${\bf \Pi},$ whichever 
is nonzero. 
For this to occur, the order of all terms in any equation in 
(\ref{originalsystemPDEs}) must be even or odd 
(as is the case in (\ref{ckdv})). 

Any term in (\ref{originalsystemPDEs}) for which the total number of 
derivatives is even contributes to the first term in (\ref{generaltypesech}); 
while any term of odd order contributes to the second term. 
Section~\ref{sechtanhmethodPDEs} deals with any case for which neither
${\bf \Gamma}$ or ${\bf \Pi}$ is identically ${\bf 0}.$ 
\vskip 2pt
\noindent
{\bf Example:} Substituting 
\begin{system}
\label{derivativesxisechPDEs}
u_{i, x_j} &=& - c_j S \sqrt{1-S^2} U_i^{\prime},  \\
u_{i, x_j x_k} &=&
c_j c_k S \sqrt{1-S^2} \left[S \sqrt{1-S^2} U_i^{\prime} \right]^{\prime} 
\\
&=& c_j c_k S [(1 - 2 S^2) U_i^{\prime} + S (1-S^2) U_i^{\prime\prime}],  \\
u_{i, x_j x_k x_l} &=& -c_j c_k c_l S \sqrt{1-S^2}
\left[S (1-2 S^2) U_i^{\prime} + 
S (1-S^2) U_i^{\prime\prime} \right]^{\prime}  \\
&=& -\!c_j c_k c_l S \sqrt{1\!-\!S^2}
\left[(1\!-\!6 S^2)U_i^{\prime}
\!+\!3S(1\!-\!2 S^2)U_i^{\prime\prime}\!+
\!S^2 (1\!-\!S^2) U_i^{\prime\prime\prime}\right], 
\end{system}
into (\ref{ckdv}), and cancelling the common $S \sqrt{1-S^2}$ factors yields
\begin{system}
\label{ckdvlegendre}
&&c_2 U_1^{\prime}\!\!-\!\!6\alpha c_1 U_1 U_1^{\prime}\!\!-\!\!\alpha c_1^3
\left[(1\!\!-\!\!6 S^2)U_1^{\prime}
\!\!+\!\! 3S (1\!\!-\!\!2S^2) U_1^{\prime\prime}
\!\!+\!\!S^2 (1\!\!-\!\!S^2) U_1^{\prime\prime\prime}\right]
\!\!+\!\!2\beta c_1 U_2 U_2^{\prime}\!\!=\!0, \\
&&c_2 U_2^{\prime}\!+\!3c_1 U_1 U_2^{\prime} \!+\!\!c_1^3 
\left[(1\!-\!6 S^2) U_2^{\prime}\!+\!3S (1\!-\!2S^2) U_2^{\prime\prime}
\!+\!S^2 (1\!-\!S^2) U_2^{\prime\prime\prime}\right]\!\!=\!0,
\end{system}
with $U_1(T)=u_1(x_1,x_2)$ and $U_2(T)=u_2(x_1,x_2).$
Note that (\ref{ckdvlegendre}) matches (\ref{legendretypesech}) with 
${\bf \Delta} = {\bf \Pi},$ since ${\bf \Gamma} = {\bf 0}.$ 
\vskip 2pt
\noindent
{\bf Step S2:$\;\;\;$Determine the degree of the polynomial solutions}
\vskip 2pt
\noindent
We seek polynomial solutions of the form,  
\begin{equation}
\label{polynomialsolutionsechPDEs}
U_i(S) = \sum_{j=0}^{M_i} a_{ij} S^j. 
\end{equation}
To determine the $M_i$ exponents, substitute $U_i (S) = S^{M_i}$ 
into the left hand side of (\ref{legendretypesech}) and proceed as 
in step T2.
Continue with step S3 for each solution of $M_i.$ 
If some of the $M_i$ exponents are undetermined, try all legitimate 
values for the free $M_i.$ 
See Section~\ref{msolve} for more details.
\vskip 2pt
\noindent
{\bf Example:} 
For (\ref{ckdv}), substituting $U_1(S) = S^{M_1}, U_2(S) = S^{M_2}$ 
into (\ref{ckdvlegendre}) and equating the highest exponents in the 
second equation yields $ M_1 + M_2 - 1 = 1 + M_2,$ or $M_1 =2.$
The maximal exponents coming from the first equation are 
$2 M_1 -1$ (from the $U_1 U_{1}^{\prime}$ term), 
$M_1 + 1$ (from $U_{1}^{\prime\prime\prime}),$
and $2 M_2 -1$ (from $U_2 U_{2}^{\prime}).$ 
Using $M_1 = 2,$ two cases emerge:
(i) the third exponent is less than the first two (equal) exponents, 
i.e., $2 M_2 -1 < 3,$ so $M_2 = 1,$ or
(ii) all three exponents are equal, in which case $M_2 = 2.$ 
For the case $M_1 = 2$ and $M_2 = 1,$
\begin{equation}
\label{ckdvchoice1} 
U(S) = a_{10} + a_{11} S + a_{12} S^2, \quad
V(S) = a_{20} + a_{21} S,
\end{equation}
and, for the case $M_1 = M_2 = 2,$ 
\begin{equation}
\label{ckdvchoice2} 
U(S) = a_{10} + a_{11} S + a_{12} S^2, \quad
V(S) = a_{20} + a_{21} S + a_{22} S^2.
\end{equation}
\vskip 1pt
\noindent
{\bf Step S3:$\;\;\;$Derive the algebraic system for the coefficients $a_{ij}$}
\vskip 2pt
\noindent 
Follow the strategy in step T3. 
\vskip 2pt
\noindent
{\bf Example:} 
After substituting (\ref{ckdvchoice1}) into (\ref{ckdvlegendre}), 
cancelling common numerical factors, and organizing the 
equations (according to complexity) one obtains
\begin{system}
\label{ckdvalgsys1}
a_{11} a_{21} c_1 &=& 0, \\
\alpha a_{11} c_1 (3 a_{12} - c_1^2) &=& 0, \\ 
\alpha a_{12} c_1 (a_{12} - 2 c_1^2) &=& 0,  \\ 
a_{21} c_1 (a_{12} - 2 c_1^2) &=& 0,  \\ 
a_{21} (3 a_{10} c_1 + c_1^3 + c_2) &=& 0,  \\
6 \alpha a_{10} a_{11} c_1 - 2 \beta a_{20} a_{21} c_1 
+ \alpha a_{11} c_1^3 - a_{12} c_2 &=& 0,  \\ 
3 \alpha a_{11}^2 c_1 + 6 \alpha a_{10} a_{12} c_1 - \beta a_{21}^2 c_1 
+ 4 \alpha a_{12} c_1^3 - a_{12} c_2 &=& 0.
\end{system}
Similarly, after substitution of (\ref{ckdvchoice2}) into 
(\ref{ckdvlegendre}), one gets
\begin{eqnarray*}
\label{ckdvalgsys2}
a_{22} c_1 (a_{12} - 4 c_1^2 ) &=& 0,  \nonumber \\ 
a_{21} ( 3 a_{10} c_1 + c_1^3 + c_2) &=& 0,  \nonumber \\ 
c_1 (a_{12} a_{21} + 2 a_{11} a_{22} - 2 a_{21} c_1^2 ) &=& 0, 
\end{eqnarray*}
\begin{eqnarray}
c_1 (3 \alpha a_{11} a_{12} -\beta a_{21} a_{22} -\alpha a_{11} c_1^2)&=& 0,
\\ 
c_1 (3 \alpha a_{12}^2 - \beta a_{22}^2 - 6 \alpha a_{12} c_1^2 ) &=& 0,
\nonumber \\ 
6 \alpha a_{10} a_{11} c_1  - 2 \beta a_{20} a_{21} c_1 
+ \alpha a_{11} c_1^3 - a_{11} c_2 &=& 0,  
\nonumber \\ 
3 a_{11} a_{21} c_1 +6 a_{10} a_{22} c_1 +8 a_{22} c_1^3 +2 a_{22} c_2 &=& 0,
\nonumber \\
3 \alpha a_{11}^2 c_1 + 6 \alpha a_{10}^2 a_{12} c_1 - \beta a_{21}^2 c_1 
- 2 \beta a_{20} a_{22} c_1 + 4 \alpha a_{12} c_1^3 - a_{12} c_2 &=& 0. 
\nonumber 
\end{eqnarray}
\vskip .005pt
\noindent
{\bf Step S4:$\;\;\;$Solve the nonlinear parameterized algebraic system}
\vskip 2pt
\noindent
Similar strategy as in step T4.
\vskip 2pt
\noindent
{\bf Example:} 
For $\alpha, \beta, c_1, c_2, a_{12} $ and $a_{21}$ all nonzero, the 
solution of (\ref{ckdvalgsys1}) is 
\begin{system}
\label{ckdvalgsol1}
a_{10} &=& -(c_1^3 + c_2)/(3 c_1), \;\;
a_{11} = 0, \;\; 
a_{12} = 2 c_1^2, \\ 
a_{20} &=& 0, \;\; 
a_{21} = \pm \sqrt{(4\alpha c_1^4 -2(1 + 2\alpha) c_1 c_2)/\beta}.
\end{system}
\noindent
For $\alpha, \beta, c_1, c_2, a_{12} $ and $a_{22}$ nonzero, 
the solution of (\ref{ckdvalgsys2}) is 
\begin{system}
\label{ckdvalgsol2}
a_{10} &=& -(4 c_1^3 + c_2)/(3 c_1), \;
a_{11} = 0, \;
a_{12} = 4 c_1^2, \\ 
a_{20} &=& \pm (4\alpha c_1^3 +(1+2\alpha)c_2)/(c_1\sqrt{6\alpha\beta}),\; 
a_{21} = 0, \; 
a_{22} = \mp 2 c_1^2 \sqrt{6\alpha / \beta}. 
\end{system}
\vskip 2pt
\noindent
{\bf Step S5:$\;\;\;$Build and test the solitary wave solutions} 
\vskip 2pt
\noindent
Substitute the result of step S4 into (\ref{polynomialsolutionsechPDEs}) 
and reverse step S1. 
Test the solutions.
\vskip 2pt
\noindent
{\bf Example:} 
The solitary wave solutions of (\ref{ckdv}) are
\begin{system}
\label{ckdvsechsol1}
u(x,t) &=& -(c_1^3 + c_2)/(3 c_1) + 2 c_1^2 
           \, {\sech}^2(c_1 x + c_2 t + \Delta), \\
v(x,t) &=& \pm \sqrt{[4 \alpha c_1^4 - 2(1 + 2\alpha) c_1 c_2]/\beta} 
           \; {\sech}(c_1 x + c_2 t + \Delta),
\end{system}
and 
\begin{system}
\label{ckdvsechsol2}
u(x,t) &\!=\!& -{(4 c_1^3 + c_2)}/{(3 c_1)} + 4 c_1^2 
           \, {\sech}^2(c_1 x + c_2 t + \Delta), \\
v(x,t) &\!=\!&
\pm (4\alpha c_1^3\!+\!(1\!+\!2\alpha)c_2)/(c_1\sqrt{6\alpha\beta})\,
\mp 2 c_1^2 \sqrt{6 \alpha/\beta}
\, {\sech}^2(c_1 x\!+\!c_2 t\!+\!\Delta).
\end{system}
In both cases $c_1, c_2, \alpha, \beta,$ and $\Delta$ are arbitrary. 
These solutions contain the solutions reported in Hereman (1991). 

Steps S1-S5 must be repeated if any of the parameters in 
(\ref{originalsystemPDEs}) are set to zero. 
% 
%%%%%%%%%%%%%Mixed tanh-sech method for nonlinear PDEs%%%%%%%%%%%%%%%%%%%%%
% 
\vspace{-0.25cm}
\section{Algorithm for mixed tanh-sech solutions for PDEs}
\label{sechtanhmethodPDEs}

The five main steps of our algorithm to compute mixed tanh-sech
solutions for (\ref{originalsystemPDEs}) are presented below.
Here we seek particular solutions of (\ref{generaltypesech}) 
when ${\bf \Gamma} \ne {\bf 0}$ and ${\bf \Pi} \ne {\bf 0}.$ 
On could apply the method of Section~\ref{sechmethodPDEs} to 
(\ref{generaltypesech}) in `squared' form
${\bf \Gamma}^2(S, {\vecU}(S), {\vecU}^{\prime}(S), \ldots ) - (1-S^2) 
\, {\bf \Pi}^2(S, {\vecU}(S), {\vecU}^{\prime}(S), \ldots ) = {\bf 0}.$
For anything but simple cases, the computations are unwieldy.
Alternatively, since $T = \sqrt{1-S^2},$ Eq.\ (\ref{generaltypesech}) may 
admit solutions of the form
\begin{equation}
\label{generalsolutionsechtanhPDEs}
U_i(S) = \sum_{j=0}^{{\tilde M}_i} \sum_{k=0}^{{\tilde N}_i} 
{\tilde a}_{i,jk} S^j T^k. 
\end{equation}
However, (\ref{generalsolutionsechtanhPDEs}) can always be rearranged 
such that
\begin{equation}
\label{polynomialsolutionsechtanhPDEs}
U_i(S) = \sum_{j=0}^{M_i} a_{ij} S^j + T\, \sum_{j=0}^{N_i} b_{ij} S^j = 
\sum_{j=0}^{M_i} a_{ij} S^j + \sqrt{1-S^2} \, \sum_{j=0}^{N_i} b_{ij} S^j.
\end{equation}
The polynomial solutions in $S$ from Section~\ref{sechmethodPDEs}
are special cases of this broader class.
Remarkably, (\ref{legendretypesech}) where $\sqrt{1-S^2}$ is not 
explicitly present also admits solutions of the form
(\ref{polynomialsolutionsechtanhPDEs}).
See Section~\ref{kdvmkdvexample} for an example. 

Computing solutions of type (\ref{polynomialsolutionsechPDEs}) with 
the tanh-sech method is inefficient and costly, 
as the following example and the examples in 
Sections~\ref{completelyintegrable} and~\ref{kdvmkdvexample} show. 
\vskip 2pt
\noindent
{\bf Example:}
We illustrate this algorithm with the system (Gao and Tian, 2001):
\begin{system}
\label{orgthreeeqs}
&& u_t - u_x - 2 v = 0, \\
&& v_t + 2 u w = 0, \\ 
&& w_t + 2 u v = 0. 
\end{system}
\vskip .005pt
\noindent
{\bf Step ST1:$\;\;\;$Transform the PDE into a nonlinear ODE}
\vskip 2pt
\noindent
Same as step S1. 
\vskip 2pt
\noindent
{\bf Example:} 
Use (\ref{chainrulesechPDE}) to transform (\ref{orgthreeeqs}) into 
\begin{system}
\label{threeeqslegendre}
&& (c_1 - c_2) S \sqrt{1-S^2}  U_1^{\prime} - 2 U_2 = 0, \\
&& c_2 S \sqrt{1-S^2} U_2^{\prime} - 2 U_1 U_3 = 0, \\
&& c_2 S \sqrt{1-S^2} U_3^{\prime} - 2 U_1 U_2 = 0. 
\end{system}
with $U_i(S)=u_i(x_1,x_2),\; i=1,2,3.$ 
\vskip 2pt
\noindent
{\bf Step ST2:$\;\;\;$Determine the degree of the polynomial solutions}
\vskip 2pt
\noindent
Seeking solutions of form (\ref{polynomialsolutionsechtanhPDEs}), 
we must first determine the leading $M_i$ and $N_i$ exponents.
Substituting $U_i (S) = a_{i0} + a_{i\, M_i} S^{M_i} + \sqrt{1-S^2} \, 
( b_{i0} + b_{i\, N_i} \, S^{N_i} )$ 
into the left hand side of (\ref{generaltypesech}), 
we get an expression of the form 
\begin{equation}
\label{PQ}
{\bf P}(S) + \sqrt{1-S^2} \, {\bf Q}(S), 
\end{equation}
where ${\bf P}$ and ${\bf Q}$ are polynomials in $S.$

Consider separately the possible balances of highest exponents in all 
$P_i$ and $Q_i.$
Then solve the resulting linear system(s) 
for the unknowns $M_i$ and $N_i.$ 
Continue with each solution in step ST3.  

In contrast to step S2, we no longer assume $M_i \geq 1, N_i \geq 1.$
Even with some $M_i$ or $N_i$ zero, non-constant solutions $U_i(S)$ 
often arise. 
In most examples, however, the sets of balance equations for 
$M_i$ and $N_i$ are too large or the corresponding linear systems are 
under-determined (i.e., several leading exponents remain arbitrary).
To circumvent the problem, 
we set all $M_i=2$ and {\rm all} $N_i=1,$ restricting the solutions to 
(at most) quadratic in $S$ and $T.$ 
\vskip 2pt
\noindent
{\bf Example:} 
For (\ref{threeeqslegendre}), 
we set all $M_i=2,\, N_i=1,$ and continue with 
\begin{equation}
\label{threeeqspols}
U_i(S) = a_{i0} + a_{i1} S + a_{i2} S^2 + \sqrt{1-S^2} \, 
(b_{i0} + b_{i1} S), \quad i = 1,2,3.
\end{equation}
\vskip .05pt
\noindent
{\bf Step ST3:$\;\;\;$Derive the algebraic system for the coefficients 
$a_{ij}$ and $b_{ij}$}
\vskip 2pt
\noindent 
Substituting (\ref{polynomialsolutionsechtanhPDEs}) into
(\ref{generaltypesech}) gives 
${\bf \widetilde{P}}(S) + \sqrt{1-S^2} \, {\bf \widetilde{Q}}(S),$ 
which must vanish identically. 
Hence, equate to zero the coefficients of the power terms in $S$ 
so that ${\bf \widetilde{P}}={\bf 0}$ and ${\bf \widetilde{Q}}={\bf 0}.$
\vskip 2pt
\noindent
{\bf Example:} After substitution of (\ref{threeeqspols}) 
into (\ref{threeeqslegendre}), the resulting nonlinear algebraic system 
for the coefficients $a_{ij}$ and $b_{ij}$ contains 25 equations (not shown).
\vskip 2pt
\noindent
{\bf Step ST4:$\;\;\;$Solve the nonlinear parameterized algebraic system}
\vskip 2pt
\noindent
In contrast to step S4 we {\rm no} longer assume that $a_{i\, M_i}$ and
$b_{i\, N_i}$ are nonzero (at the cost of generating some constant solutions, 
which we discard later). 
\vskip 2pt
\noindent
{\bf Example:} For (\ref{orgthreeeqs}), there are 11 solutions. 
Three are trivial, leading to constant $U_i.$
Eight are nontrivial solutions giving the results below. 
\vskip 2pt
\noindent
{\bf Step ST5:$\;\;\;$Build and test the solitary wave solutions} 
\vskip 2pt
\noindent
Proceed as in step S5. 
\vskip 2pt
\noindent
{\bf Example:} The solitary wave solutions of (\ref{orgthreeeqs}) are:
\begin{system}
\label{threeeqssolution1}
u(x,t) &=& \pm c_2 \, {\tanh}\,\xi,  \\
v(x,t) &=& \mp \textstyle{\frac{1}{2}} c_2 (c_1 - c_2) \, {\sech}^2\xi, \\ 
w(x,t) &=& - \textstyle{\frac{1}{2}} c_2 (c_1 - c_2) \, {\sech}^2\xi, 
\end{system}
which could have been obtained with the tanh-method 
of Section~\ref{tanhmethodPDEs};
\begin{system}
\label{threeeqssolution2}
u(x,t) &=& \pm i c_2 \, {\sech}\,\xi,  \\
v(x,t) &=& \pm \textstyle{\frac{1}{2}} i c_2 (c_1 - c_2) \, 
{\tanh}\,\xi \, {\sech}\,\xi, \\ 
w(x,t) &=& \textstyle{\frac{1}{4}} c_2 (c_1 - c_2) \, 
\left( 1 - 2 \, {\sech}^2\xi \right),  
\end{system}
reported in Gao and Tian (2001); and the two complex solutions 
\begin{system}
\label{threeeqssolution3}
u(x,t) &=& \pm \textstyle{\frac{1}{2}} i c_2 \,
\left( {\sech}\,\xi \pm i \, {\tanh}\,\xi \right), \\
v(x,t) &=& \textstyle{\frac{1}{4}} c_2 (c_1 - c_2) 
\, {\sech}\,\xi \left( {\sech}\,\xi \pm i \, {\tanh}\,\xi \right), \\ 
w(x,t) &=& - \textstyle{\frac{1}{4}} c_2 (c_1 - c_2) 
\, {\sech}\,\xi \left( {\sech}\,\xi \pm i \, {\tanh}\,\xi \right). 
\end{system}
In all solutions $\xi = c_1 x + c_2 t + \Delta,$ with $c_1, c_2$ 
and $\Delta$ arbitrary.
The complex conjugates of (\ref{threeeqssolution3}) are also solutions. 
% 
%%%%%%%%%%%%%%%%%Cn method for nonlinear PDEs%%%%%%%%%%%%%%%%%%%%%%
% 
\section{Algorithms to compute sn and cn solutions for PDEs}
\label{cnmethodPDEs}
In this section we give the main steps (labelled \mbox{CN1}-\mbox{CN5}) 
of our algorithm to compute polynomial solutions of (\ref{originalsystemPDEs}) 
in terms of Jacobi's elliptic cosine function (cn). 
Modifications needed for solutions involving the $\sn$ function are given
at the end of this section.
Details for steps \mbox{CN2}, \mbox{CN4} and \mbox{CN5} are shown in 
Section~\ref{algorithms}.
\vskip 2pt
\noindent
{\bf Example:} Consider the KdV equation (Ablowitz and Clarkson, 1991),
\begin{equation}
\label{orgkdv}
u_t + \alpha u u_x + u_{xxx} = 0,
\end{equation}
with real constant $\alpha.$ 
The KdV equation models, among other things, waves in shallow water
and ion-acoustic waves in plasmas.
\vskip 2pt
\noindent
{\bf Step CN1:$\;\;\;$Transform the PDE into a nonlinear ODE}
\vskip 2pt
\noindent
Similar to the strategy in T1 and S1, using (Lawden, 1989)
\begin{equation}
{\sn}^2(\xi;m) = 1 - {\cn}^2(\xi;m), \;\;
{\dn}^2(\xi;m) = 1 - m + m \, {\cn}^2(\xi;m),
\end{equation}
and 
\begin{equation}
\label{cnderivative}
{\cn}^{\prime}(\xi;m) = - {\sn}(\xi;m) {\dn}(\xi;m), 
\end{equation}
one has $\mbox{CN}^{\prime} = 
-\sqrt{(1 - \mbox{CN}^2)(1 - m + m\,\mbox{CN}^2)}$ 
where $\mbox{CN} = {\cn}(\xi;m)$ is the Jacobi elliptic cosine with 
argument $\xi$ and modulus $0\leq m \leq 1$.

Repeatedly applying the chain rule
\begin{equation}
\label{chainrulecnPDE}
\frac{\partial \bullet}{\partial x_j} 
= \frac{d\bullet}{d\mbox{CN}} 
\frac{d\mbox{CN}}{d\xi} \frac{\partial \xi}{\partial x_j} 
= - c_j \sqrt{(1 - 
\mbox{CN}^2)(1 - m + m\,\mbox{CN}^2)} \frac{d\bullet}{d\mbox{CN}}, 
\end{equation}
system (\ref{originalsystemPDEs}) is transformed into a nonlinear ODE system.
In addition to the $c_j,$ the algorithm introduces $m$ as an extra parameter.
\vskip 2pt
\noindent
{\bf Example:} Using (\ref{chainrulecnPDE}) to transform (\ref{orgkdv}) 
we have
\begin{equation}
\label{kdvCNODE}
\begin{array}{cl}
&\!\!\!\left(  c_1^3 ( 1 - 2 m + 6 m \,\mbox{CN}^2 ) - 
c_2 - \alpha c_1 U_1 \right) U_1^{\prime} 
\nonumber \\
+\!\!&\! 3 c_1^3 \mbox{CN} (1 - 2 m + 2 m\,\mbox{CN}^2) U_1^{\prime\prime} 
-\!\! c_1^3 (1 \!- \mbox{CN}^2)(1 - m + m\,\mbox{CN}^2 ) 
U_1^{\prime\prime\prime} = 0
\end{array}
\end{equation}
{\bf Step CN2:$\;\;\;$Determine the degree of the polynomial solutions}
\vskip 2pt
\noindent
Follow the strategy in step T2.
\vskip 2pt
\noindent
{\bf Example:} For (\ref{orgkdv}), substituting 
$U_1(\mbox{CN}) = \mbox{CN}^{M_1}$ into (\ref{kdvCNODE}) and equating the 
highest exponents gives
$ 1 + M_1 = -1 + 2\,M_1.$
Then, $M_1 = 2,$ and
\begin{equation}
U_1(\mbox{CN}) = a_{10} + a_{11}\mbox{CN} + a_{12}\mbox{CN}^2.
\label{kdvCNpols}
\end{equation}
\vskip 0.005pt
\noindent
{\bf Step CN3:$\;\;\;$Derive the algebraic system for the 
coefficients $a_{ij}$}
\vskip 2pt
\noindent
Proceed as in step T3.
\vskip 2pt
\noindent
{\bf Example:} For (\ref{orgkdv}), after substituting 
(\ref{kdvCNpols}) into (\ref{kdvCNODE}), one finds
\begin{system}
\label{kdvCNsystem}
a_{11}\,c_1\,(\alpha\,a_{12} - 2\,m\,c_1^2) &=& 0, \\
a_{12}\,c_1\,(\alpha\,a_{12} - 12\,m\,c_1^2) &=& 0,  \\
a_{11}\,(\alpha\,a_{10}\,c_1 - c_1^3 + 2\,m\,c_1^3 + c_2) &=& 0,  \\
\alpha\, a_{11}^2\,c_1 + a_{12}\,( 2\,\alpha\,a_{10}\, c_1
- 16 m\, c_1^3 - 8\, c_1^3 + 2\, c_2 )&=& 0. 
\end{system}
\vskip 0.005pt
\noindent
{\bf Step CN4:$\;\;\;$Solve the nonlinear parameterized algebraic system}
\vskip 2pt
\noindent
Solve the system as in step T4.
\vskip 2pt
\noindent
{\bf Example:} For $c_1, c_2, m, \alpha $ and $a_{12}$ nonzero, 
the solution of (\ref{kdvCNsystem}) is
\begin{equation}
a_{10}  =  [4 c_1^3 \, (1 - 2 \,m)- c_2] / (\alpha \,c_1), \;
a_{11}  =  0, \;
a_{12}  =  (12 \,m \, c_1^2)/ \alpha.
\end{equation}
\vskip 0.005pt
\noindent
{\bf Step CN5:$\;\;\;$Build and test the solitary wave solutions}
\vskip 2pt
\noindent
Substitute the results of step CN4 into (\ref{kdvCNpols}).
Reverse step CN1.
Test the solutions.
\vskip 2pt
\noindent
{\bf Example:} The cnoidal wave solution of (\ref{orgkdv}) is
\begin{equation}
u(x,t) = {[4 c_1^3 (1 - 2m) - c_2]}/{(\alpha c_1)} +
{(12 m\, c_1^2)}/{(\alpha)} \, {\cn}^2(c_1 x + c_2 t + \Delta; m),  
\label{kdvCNsoln}
\end{equation}
where $c_1, c_2, \alpha, \Delta$ and modulus $m$ are arbitrary. 
If any of the parameters in (\ref{originalsystemPDEs}) are zero, 
steps CN1-CN5 should be repeated. 
\vskip 2pt
\noindent
{\bf Computation of solutions involving Sn}
\vskip 2pt
\noindent
To find solutions in terms of Jacobi's $\sn$ function, one uses the identities,
\begin{system}
\label{dncnidentity2}
&& {\cn}^2(\xi;m) = 1 - {\sn}^2(\xi;m), \;\;
{\dn}^2(\xi;m) = 1 - m \, {\sn}^2(\xi;m), \\
&& {\sn}^{\prime}(\xi;m)  = {\cn}(\xi;m) \, {\dn}(\xi;m).
\end{system}
Then, $\mbox{SN}^{\prime} = 
\sqrt{(1 - \mbox{SN}^2)(1 - m\,\mbox{SN}^2)},$ where
$\mbox{SN} = {\sn}(\xi;m)$ is the Jacobi elliptic sine with argument $\xi$ 
and modulus $0 \leq m \leq 1.$
The steps are identical to the $\cn$ case, except one uses the chain rule  
\begin{equation}
\label{chainrulesnPDE}
\frac{\partial \bullet}{\partial x_j} 
= \frac{d\bullet}{d\mbox{SN}} \frac{d\mbox{SN}}{d\xi} \frac{\partial 
\xi}{\partial x_j} 
= c_j \sqrt{(1-\mbox{SN}^2)(1 - m\, 
\mbox{SN}^2)} \frac{d \bullet}{d \mbox{SN}}.
\end{equation}
Since (\ref{chainrulecnPDE}) and (\ref{chainrulesnPDE}) involve roots, 
as in Sections~\ref{sechmethodPDEs} and~\ref{sechtanhmethodPDEs} 
there is no reason to restrict the solutions to polynomials in 
only $\cn$ or $\sn.$ 
Solutions involving both $\sn$ and $\cn$ 
(or combinations with $\dn$) are beyond the scope of this paper. 

Finally, from the $\sn$ and $\cn$ solutions, $\sin, \cos, \sech,$ 
and tanh-solutions can be obtained by taking the appropriate limits 
for the modulus $(m \rightarrow 0,$ and $m \rightarrow 1).$ 
Indeed, 
$ \sn(\xi;0) \!=\! \sin(\xi), \, \sn(\xi;1) \!=\! \tanh(\xi), \,
 \cn(\xi;0) \!=\! \cos(\xi), \, \cn(\xi;1) \!=\! \sech(\xi).$
No need to compute solutions in $\dn$ explicitly since 
$\cn(\sqrt{m}\xi;1/m)= \dn(\xi;m).$
% 
%%%%%%%%%%%%%%%%%%Algorithm%%%%%%%%%%%%%%%%%%%%%%%%%%%%
% 
\vspace{-0.25cm}
\section{Key Algorithms}
\label{algorithms}
In this section we present in a uniform manner the details of steps two, four 
and five of the algorithms in Sections~\ref{tanhmethodPDEs}-\ref{cnmethodPDEs}.
\vspace{-0.50cm}
\subsection{Algorithm to compute the degree of the polynomials}
\label{msolve}
\vskip 0.001pt
\noindent
\vspace{-0.11cm}
\noindent
{\bf Step M1:$\;\;\;$Substitute the leading-order ansatz}
\vskip 2pt 
\noindent 
A tracking variable is attached to each term in the original system of PDEs. 
Let \verb|Tr[i]| denote the tracking variable of the $i$th term in 
(\ref{originalsystemPDEs}).

The first step of the main algorithms 
leads to a system of parameterized ODEs in $\myvec{U,U',U'',\dots,U}^{(m)}.$ 
These ODEs match the form
\begin{equation}
 \label{PQf}
 \myvec{\Gamma}(F, {\vecU}(F), {\vecU}^{\prime}(F), \ldots ) + 
 \sqrt{R(F)} \, \myvec{\Pi}(F, {\vecU}(F), {\vecU}^{\prime}(F), \ldots )
= \myvec{0},
\end{equation}
where $F$ is either $T,S,$ \mbox{CN}, or \mbox{SN}, and $R(F)$ is defined 
in Table \ref{tbl:R(F)}.
\vskip 0.001pt
\noindent
\begin{table}
$$ \begin{array}{c|c}
 F & R(F) \\ \hline \hline
 T & 0 \\ \hline
 S & 1-S^2 \\ \hline
 \mbox{CN} & (1 - \mbox{CN}^2)(1 - m + m\,\mbox{CN}^2) \\ \hline
 \mbox{SN} & (1 - \mbox{SN}^2)(1 - m\,\mbox{SN}^2)
\end{array}$$
\vspace{-0.50cm}
\caption{Values for $R(F)$ in (\ref{PQf}).}
\label{tbl:R(F)}
\vspace{-0.65cm}
\end{table}
\vskip 0.0001pt
\noindent 
Since the highest degree term only depends on $F^{M_i},$ 
it suffices to substitute
\begin{equation}
\label{ansatz}
U_i(F) \to F^{M_i}
\end{equation}
into (\ref{PQf}).
\begin{example} 
We use the coupled KdV equations (\ref{orgckdv}) as our leading example:
\begin{system}
\label{ckdv:repckdv}
\verb|Tr[1]| u_t 
- 6 \alpha \verb|Tr[2]| u u_x + 2 \beta \verb|Tr[3]| v v_x 
- \alpha \verb|Tr[4]| u_{xxx} &=& 0,  \\
\verb|Tr[5]| v_t + 3 \verb|Tr[6]| u v_x 
+ \verb|Tr[7]| v_{xxx} &=& 0.
\end{system}
Step S1 resulted in (\ref{ckdvlegendre}) with $\myvec{\Pi} = \myvec{0}$. 
Substituting (\ref{ansatz}) into (\ref{ckdv:repckdv}), we get
\begin{system}
\label{ckdv:ansatz}
&& ( \verb|Tr[1]| c_2 M_1 - \alpha \verb|Tr[4]| c_1^3 M_1^3) S^{M_1 - 1} 
+\alpha \verb|Tr[4]| c_1^3 M_1 (M_1 + 1) (M_1 + 2) S^{M_1 + 1} \\
&&  \qquad \quad 
- 6 \alpha \verb|Tr[2]| c_1 M_1 S^{2M_1 - 1}  
+ 2 \beta \verb|Tr[3]| c_1 M_2 S^{2 M_2 - 1}  = 0, \\
&& (\verb|Tr[5]| c_2 M_2 + \verb|Tr[7]| c_1^3 M_2^3 ) S^{M_2 - 1}
- \verb|Tr[7]| c_1^3 M_2 (M_2 + 1 ) (M_2 + 2 ) S^{M_2 + 1} \\ 
&&  \qquad \quad 
+ 3 \verb|Tr[6]| c_1 M_2 S^{M_1 + M_2 - 1} = 0.
\end{system}
\end{example}
\vskip 0.001pt
\noindent
\vspace{-0.11cm}
\noindent
{\bf Step M2:$\;\;\;$Collect exponents and prune sub-dominant branches}
\vskip 2pt
\noindent 
The balance of highest exponents must come from different terms in 
(\ref{originalsystemPDEs}).  
For each equation $\Delta_i$ and for each tracking variable, 
collect the exponents of $F,$ remove duplicates, and non-maximal exponents.
For example, $M_1 - 1$ can be removed from $\{M_1 + 1, M_1 - 1\}$ because 
$M_1 + 1 > M_1 - 1$.  
\vskip 0.01pt
\begin{example}
Collecting the exponents of $S$ in (\ref{ckdv:ansatz}), we get the 
unpruned list:
\begin{equation}
\label{ckdv:exponents}
   \begin{array}{rl|rl} 
   \multicolumn{2}{c}{\Delta_1} & \multicolumn{2}{c}{\Delta_2} \\ \hline
     \!\!\!\verb|Tr[1]|\!:\!\!& 
     \{ M_1 \!-\! 1\} & \!\verb|Tr[5]|\!:\!\! & \{ M_2 \!-\! 1 \} \\  
     \!\!\!\verb|Tr[2]|\!:\!\!& 
     \{ 2M_1 \!-\! 1\}& \!\verb|Tr[6]|\!:\!\! & \{ M_1 \!+\! M_2 \!-\! 1 \} \\
     \!\!\!\verb|Tr[3]|\!:\!\!& 
     \{ 2M_2 \!-\! 1\} &\!\verb|Tr[7]|\!:\!\! & \{ M_2 \!+\! 1, M_2 \!+\! 1, 
     M_2 \!+\! 1, M_2 \!-\! 1 \} \\
     \!\!\!\verb|Tr[4]|\!:\!\!& 
     \{ M_1 \!+\! 1, M_1 \!+\! 1, M_1 \!+\! 1, M_1 \!-\! 1 \}
   \end{array}
\end{equation}
\vskip .001pt
\noindent
We prune by removing duplicates and non-maximal expressions, and get
\begin{equation}
 \label{ckdv:Mis}
 \begin{array}{rl}
  {\rm from} \; \Delta_1 : & \{ M_1 + 1, 2M_1 - 1, 2M_2 - 1 \}, \\
  {\rm from} \; \Delta_2 : & \{ M_2 + 1, M_1 + M_2 -1 \}.
 \end{array}
\end{equation}
\end{example}
\vskip 0.001pt
\noindent
{\bf Step M3:$\;\;\;$Combine expressions and compute relations for $M_i$}
\vskip 2pt
\noindent 
For each $\Delta_i$ separately, equate all possible combinations of two 
elements. 
Construct relations between the $M_i$ by solving for one $M_i.$ 
\vskip 0.01pt
\noindent
\begin{example}
Combining the expressions in (\ref{ckdv:Mis}), we get 
\begin{equation}
 \begin{array}{r@{\,=\,}l|r@{\,=\,}l} 
  \multicolumn{2}{c}{\Delta_1} & \multicolumn{2}{c}{\Delta_2} \\ \hline
  M_1 + 1 & 2 M_1 - 1 & M_2 + 1 & M_1 + M_2 - 1 \\
  M_1 + 1 & 2 M_2 - 1 \\
  2 M_1 - 1 & 2 M_2 - 1 \\
 \end{array}
\end{equation}
We construct relations between the $M_i$ by solving for $M_1$ (in this case): 
\begin{equation}
 \begin{array}{r@{\,=\,}l|r@{\,=\,}l}
  \multicolumn{2}{c}{\Delta_1} & \multicolumn{2}{c}{\Delta_2} \\ \hline
  M_1 & 2 & M_1 & 2 \\
  M_1 & 2 M_2 - 2 \\
  M_1 & M_2 
 \end{array}
\end{equation}
\end{example}
\vskip 0.001pt
\noindent
\vspace{-0.11cm}
\noindent
{\bf Step M4:$\;\;\;$Combine relations and solve for exponents $M_i$}
\vskip 2pt 
\noindent 
By combining the lists of expressions in an outer product like fashion, 
we generate all the possible linear equations for $M_i.$  
Solving this linear system, we form a list of all the possible solutions 
for $M_i.$
\vskip 0.01pt
\noindent
\begin{example}
Combining the equations in $\Delta_1$ and $\Delta_2$, we obtain
\begin{equation}
\{M_1 = 2, M_1 = 2\}, \;
    \{M_1 = 2, M_1 = M_2\}, \;
   \{M_1 = 2, M_1 = 2 M_2 - 2\}.
\end{equation}
Solving, we find
\begin{equation}
 \label{ckdv:mSoln1}
 \left\{
 \begin{array}{l}
   M_1 = 2 \\
   M_2 = 2
 \end{array}
 \right. \qquad
 \left\{
 \begin{array}{l}
   M_1 = 2 \\
   M_2 = \rm Free
 \end{array}
 \right.
\end{equation}
\end{example}
\vskip 0.001pt
\noindent
\vspace{-0.11cm}
\noindent
{\bf Step M5:$\;\;\;$Discard invalid exponents $M_i$}
\vskip 2pt
\noindent
The solutions are substituted into the unpruned list of exponents 
(in step M2). 
For every solution (without free exponents) we test whether or not there is 
a highest-power balance between at least two different tracking variables.  
If not, the solution is rejected.
Non-positive, fractional, and complex exponents are discarded 
(after showing them to the user). 
Negative exponents $(M_i = -p_i)$ and fractional exponents 
($M_i={p_i}/{q_i})$ indicate that a change of dependent variables 
($u_1 = \tilde{u_i}^{-p_i}$ or $u_i = \tilde{u_i}^{{1}/{q_i}})$ 
should be attempted in (\ref{originalsystemPDEs}). 
Presently, such nonlinear transformations are only carried out automatically
for single equations.
\vskip 0.01pt
\noindent
\begin{example}
Removing the case $\{M_1 = 2, M_2 = \rm Free\}$ from (\ref{ckdv:mSoln1}), 
we substitute $\{M_1 = 2, M_2 = 2\}$ into (\ref{ckdv:exponents}). 
Leading exponent $(3$ in this case) occurs for 
\verb|Tr[2]|, \verb|Tr[3]| and \verb|Tr[4]| in $\Delta_1,$ 
and for \verb|Tr[6]| and \verb|Tr[7]| in $\Delta_2.$ 
The solution is accepted.
\end{example}
\vskip 0.001pt
\noindent
\vspace{-0.11cm}
\noindent
{\bf Step M6:$\;\;\;$Fix undetermined $M_i$ and generate additional solutions}
\vskip 2pt
\noindent
When some solutions involve one or more arbitrary $M_i$ we produce 
candidate solutions with a countdown procedure and later 
reject invalid candidates.

Based on the outcome of step M5, scan for freedom in one or more of 
$M_i$ by gathering the highest-exponent expressions from the unpruned list 
in step M2.  
If the dominant expressions are free of any of the $M_i$, a countdown 
mechanism generates valid integer values for those $M_i.$ 
These values of $M_i$ must not exceed those computed in step M5.  
Candidate solutions are tested (and rejected, if necessary) by the 
procedure in step M5.  
\vskip 0.01pt
\noindent
\begin{example}
The dominant expressions from (\ref{ckdv:exponents}) with 
$\{M_1 = 2, M_2 = 2\}$ are
\begin{equation}
  \begin{array}{rl|rl} 
  \multicolumn{2}{c}{\Delta_1} & \multicolumn{2}{c}{\Delta_2} \\ \hline
    \verb|Tr[2]|: & \{2 M_1 - 1\} & \verb|Tr[6]|: & \{M_1 + M_2 - 1\} \\
    \verb|Tr[3]|: & \{2 M_2 - 1\} & \verb|Tr[7]|: &  \{M_2 + 1\} \\
    \verb|Tr[4]|: & \{M_1 + 1 \} 
  \end{array}
\end{equation}
Substituting $M_1 = 2$, the highest exponent $(3$ in this case) matches for 
\verb|Tr[2]| and \verb|Tr[4]| in $\Delta_1$ when $M_2 \leq 2.$   
The highest exponent $(M_2+1)$ matches for \verb|Tr[6]| and 
\verb|Tr[7]| in $\Delta_2.$

A countdown mechanism then generates the following list of candidates 
\begin{equation}
 \left\{
 \begin{array}{l}
   M_1 = 1 \\
   M_2 = 1
 \end{array}
 \right. \qquad
 \left\{
 \begin{array}{l}
   M_1 = 1 \\
   M_2 = 2
 \end{array}
 \right. \qquad
 \left\{
 \begin{array}{l}
   M_1 = 2 \\
   M_2 = 1
 \end{array}
 \right. \qquad
 \left\{
 \begin{array}{l}
   M_1 = 2 \\
   M_2 = 2
 \end{array}
 \right.
\end{equation}
Verifying these candidate solutions, we are left with 
\begin{equation}
 \left\{
 \begin{array}{l}
   M_1 = 2 \\
   M_2 = 1
 \end{array}
 \right. \qquad
 \left\{
 \begin{array}{l}
   M_1 = 2 \\
   M_2 = 2
 \end{array}
 \right.
\end{equation}
\end{example}
Notice that for the new solution $\{ M_1=2, M_2=1 \} $ only the exponents
corresponding to \verb|Tr[2]| and \verb|Tr[4]| in $\Delta_1$ are equal.

Currently, for the mixed tanh-sech method, the code sets 
$M_i = 2$ and $N_i = 1.$
\vspace{-0.50cm}
\subsection{Algorithm to analyze and solve nonlinear algebraic systems}
\label{analyzeandsolve}

In this section, we detail our algorithm to analyze and solve
nonlinear parameterized algebraic systems (as generated in step 3 of
the main algorithms).  
Our solver is custom-designed for systems that are (initially) polynomial 
in the primary unknowns ($a_{ij}$), the secondary unknowns ($c_i$), and 
parameters $(m, \alpha,\beta,\gamma,\dots ).$

The goal is to compute the coefficients $a_{ij}$ in terms of the wave 
numbers $c_i$ and the parameters $m, \alpha, \beta,$ etc. 
In turn, the $c_i$ must be solved in terms of these parameters.
Possible compatibility conditions for the parameters (relations amongst
them or specific values for them) must be added to the solutions.

Algebraic systems are solved recursively, starting with the simplest equation,
and continually back-substituting solutions.
This process is repeated until the system is completely solved.

To guide the recursive process, we designed functions to
(i)   factor, split, and simplify the equations; 
(ii)  sort the equations according to their complexity; 
(iii) solve the equations for sorted unknowns; 
(iv)  substitute solutions into the remaining equations; and
(v)   collect the solution branches and constraints.

This strategy is similar to what one would do by hand. 
If there are numerous parameters in the system or if it is of high degree, 
there is no guarantee that our solver will return a suitable result, 
let alone a complete result. 
\vskip 0.001pt
\noindent
\vspace{-0.11cm}
\noindent
{\bf Step R1:$\;\;\;$Split and simplify each equation}
\vskip 2pt
\noindent
For all but the mixed tanh-sech algorithm, we assume that 
the coefficients $a_{iM_i}$ of the highest power terms are nonzero 
and that $c_i,m,\alpha,\beta,$ etc.\ are nonzero.
For the mixed sech-tanh method, $a_{iM_i}=a_{i2}$ and 
$b_{iN_i}=b_{i1} $ are allowed to be zero. 

We first factor equations and set admissible factors equal to zero 
(after clearing possible exponents). 
For example, $\{\phi_1{\phi_2}^3{\phi_3}^2 \!=\! 0\} \to 
\{\phi_1 \!=\! 0, \phi_2 \!=\! 0, \phi_3 \!=\! 0\},$ 
where $\phi_i$ is a polynomial in primary and secondary unknowns along 
with the parameters.  
Equations where non-zero expressions are set to zero are disgarded.
\vskip 0.01pt
\noindent
\begin{example} 
Consider (\ref{ckdvalgsys2}), which was derived in the search for 
sech-solutions of (\ref{orgckdv}) for case $M_1=M_2=2.$
Taking $a_{12}, a_{22}, c_1, c_2, \alpha, \beta,$ to be nonzero, 
splitting equations, and removing non-zero factors leads to
\begin{system} 
  \label{ckdvr1}
  a_{12} - 4 c_1^2 &=& 0, \\
  a_{21} = 0 \lor (3 a_{10} c_1 + c_1^3 + c_2) &=& 0, \\
  a_{12} a_{21} + 2 a_{11} a_{22} - 2 a_{21} c_1^2 &=& 0, \\ 
  3 \alpha a_{11} a_{12} - \beta a_{21} a_{22} - \alpha a_{11} c_1^2 &=& 0, \\ 
  3 \alpha a_{12}^2 - \beta a_{22}^2 - 6 \alpha a_{12} c_1^2 &=& 0, \\ 
  6 \alpha a_{10} a_{11} c_1 -2 \beta a_{20} a_{21} c_1 + \alpha a_{11} c_1^3 
  - a_{11} c_2 &=& 0, \\ 
  3 a_{11} a_{21} c_1 + 6 a_{10} a_{22} c_1 + 8 a_{22} c_1^3 + 2 a_{22} c_2 
  &=& 0, \\ 
  3 \alpha a_{11}^2 c_1 + 6 \alpha a_{10} a_{12} c_1 - \beta a_{21}^2 c_1 
  - 2 \beta a_{20} a_{22} c_1 + 4 \alpha a_{12} c_1^3 - a_{12} c_2 &=& 0, 
\end{system}
where $\lor$ is the logical or. 
\end{example}
\vskip 0.001pt
\noindent
\vspace{-0.11cm}
\noindent
{\bf Step R2:$\;\;\;$Sort equations according to complexity}
\vskip 2pt
\noindent
A heuristic measure of complexity is assigned to each $\phi_i$ by 
computing a weighted sum of the degrees of nonlinearity in the primary 
and secondary unknowns, parameters, and the length of $\phi_i.$
Linear and quasi-linear equations (with products like 
$a_{11}a_{21})$ are of lower complexity than polynomial equations
of higher degree or non-polynomial equations.
Solving the equation of the lowest complexity first, forestalls branching, 
avoids expression swell, and conserves memory.
\vskip 0.01pt
\noindent
\begin{example}
Sorting (\ref{ckdvr1}), we get
\begin{system} 
a_{12} - 4 c_1^2 &=& 0, \\
3 \alpha a_{11} a_{12} - \beta a_{21} a_{22} - \alpha a_{11} c_1^2 &=& 0, \\
a_{12} a_{21} + 2 a_{11} a_{22} - 2 a_{21} c_1^2 &=& 0, \\ 
a_{21} = 0 \lor (3 a_{10} c_1 + c_1^3 + c_2) &=& 0, \\
3 \alpha a_{12}^2 - \beta a_{22}^2 - 6 \alpha a_{12} c_1^2 &=& 0, \\ 
6 \alpha a_{10} a_{11} c_1 -2 \beta a_{20} a_{21} c_1 + \alpha a_{11} c_1^3 
- a_{11} c_2 &=& 0, \\ 
3 a_{11} a_{21} c_1 + 6 a_{10} a_{22} c_1 + 8 a_{22} c_1^3 + 2 a_{22} c_2 
&=& 0, \\ 
3 \alpha a_{11}^2 c_1 + 6 \alpha a_{10} a_{12} c_1 - \beta a_{21}^2 c_1 
- 2 \beta a_{20} a_{22} c_1 + 4 \alpha a_{12} c_1^3 - a_{12} c_2 &=& 0.
\end{system}
\end{example}
\vskip 0.001pt
\noindent
\vspace{-0.11cm}
\noindent
{\bf Step R3:$\;\;\;$Solve equations for ordered unknowns}
\vskip 3pt
\noindent
The ordering of unknowns is of paramount importance.
The unknowns from the first equation from step R2 are ordered so that the 
lowest exponent primary-unknowns precede the primary-unknowns that the 
equation is not polynomial in.  
If there are not any primary-unknowns, the lowest exponent 
secondary-unknowns precede the secondary-unknowns that the 
equation is not polynomial in.  
Likewise, in the absence of primary- or secondary-unknowns, the lowest 
exponent parameters precede the non-polynomial parameters.  

The equation is solved using the built-in {\it Mathematica} function 
\verb|Reduce|, which produces a list of solutions and constraints.
Constraints of the form $a \neq b$ (where neither $a$ or $b$ is zero) are 
pruned, and the remaining constraints and solutions are collected.  
\vskip 0.1pt
\noindent
\begin{example}
In this example, $a_{12} - 4c_1^2 = 0$ is solved for $a_{12}$ and the solution 
$a_{12} = 4 c_1^2$ is added to a list of solutions.
\end{example}
\vskip 0.001pt
\noindent
\vspace{-0.08cm}
\noindent
{\bf Step R4:$\;\;\;$Recursively solve the entire system}
\vskip 2pt
\noindent
The solutions and constraints from step R3 are applied and added to the 
previously found solutions and constraints.  
In turn, all the solutions are then applied to the remaining equations.  
The updated system is simplified by clearing common denominators in each
equation and continuing with the numerators.  
Steps R1-R4 are then repeated on the simplified system.  
\vskip 0.01pt
\noindent
\begin{example}
  Substituting $a_{12} = 4 c_1^2$ and clearing denominators, we obtain
  \begin{system}
  \beta a_{21} a_{22} - 11 \alpha a_{11} c_1^2 &=& 0 \\ 
  a_{11} a_{22} + a_{21} c_1^2 &=& 0, \\ 
  a_{21} = 0 \lor (3 a_{10} c_1 + c_1^3 + c_2) &=& 0, \\ 
  \beta a_{22}^2 - 24 \alpha c_1^4 &=& 0, \\ 
  6 \alpha a_{10} a_{11} c_1 - 2 \beta a_{20} a_{21} c_1 + \alpha a_{11} c_1^3 
  - a_{11} c_2 &=& 0, \\ 
  3 a_{11} a_{21} c_1 + 6 a_{10} a_{22} c_1 + 8 a_{22} c_1^3 + 2 a_{22} c_2 
  &=& 0, \\ 
  3 \alpha a_{11}^2 - \beta a_{21}^2 - 2 \beta a_{20} a_{22} + 24 \alpha 
  a_{10} c_1^2 + 16 \alpha c_1^4 - 4 c_1 c_2 &=& 0. 
  \end{system}
The recursive process terminates when the system is completely solved. 
The solutions (including possible constraints) are returned.

Repeating steps R1-R4 seven more times the {\em global} solution of 
(\ref{ckdvalgsys2}) is obtained:
\begin{system}
\label{repckdvalgsol2}
a_{10} &=& -(4 c_1^3 + c_2)/(3 c_1), \;
a_{11} = 0, \;
a_{12} = 4 c_1^2, \\ 
a_{20} &=& \pm (4\alpha c_1^3 +(1+2\alpha)c_2)/(c_1\sqrt{6\alpha\beta}),\; 
a_{21} = 0, \; 
a_{22} = \mp 2 c_1^2 \sqrt{6\alpha / \beta}. 
\end{system}
where $c_1, c_2, \alpha $ and $ \beta$ are arbitrary. 

This solution of (\ref{ckdvalgsys1}), corresponds to the  
$M_1 = 2, M_2 = 1$ case given in (\ref{ckdvalgsol1}).
\end{example}
\vspace{-0.50cm}
\subsection{Algorithm to build and test solutions}
\label{testing}

The solutions to the algebraic system found in 
Section~\ref{analyzeandsolve} are substituted into 
\begin{equation}
u_i(\vecx) = 
\sum_{j=0}^{M_i} a_{ij} F^j(\xi) + 
\sqrt{R(F)} \sum_{j=0}^{N_i} b_{ij} F^j(\xi),
\end{equation}
where $F$ and $R(F)$ are defined in Section~\ref{msolve}.  
The constraints on the parameters ($m,\alpha,\beta,$ etc.) are also 
collected and applied to system (\ref{originalsystemPDEs}).

Since the algorithm used to solve the nonlinear algebraic system continually 
clears denominators, it is important to test the final solutions for $u_i.$  
While {\it Mathematica's} \verb|Reduce| function generates constraints that 
should prevent any undetermined or infinite coefficients $a_{ij}$ after 
back-substitution, it is still prudent to check the solutions.  

To present solutions in the simplest format, we assume that all parameters 
$(c_i, m, \alpha, \beta,$ etc.) are positive, real numbers. 
This allows us to repeatedly apply rules like
$\sqrt{\alpha^2} \rightarrow \alpha,$
$\sqrt{-\alpha^2} \rightarrow i\,\alpha,$  
$\sqrt{-\beta} \rightarrow i\,\sqrt{\beta}$
and $\sqrt{-(c_1 + c_2)^2} \rightarrow i\,(c_1+c_2).$

We allow for two flavors of testing: a numeric test for complicated 
solutions and a symbolic test which guarantees the solution.
In either test, we substitute the solutions into (\ref{originalsystemPDEs}) 
after casting the solutions into exponential form, 
i.e., $\tanh\xi \to (e^\xi - e^{-\xi})/(e^\xi + e^{-\xi})$ 
and $\sech\xi \to 2/(e^\xi + e^{-\xi}).$  
\vskip 1pt
\noindent
For the numeric test of solutions:
\begin{itemize}
 \item after substituting the solution, substitute random real numbers 
 in $[-1,1]$ for $x_i, c_i,$ and $\Delta$ in the left hand side of 
 (\ref{originalsystemPDEs}),
 \item expand and factor the remaining expressions,
 \item substitute random real numbers in $[-1,1]$ for arbitrary
 $a_{ij}, b_{ij}, m, \alpha, \dots,$
 \item expand and factor the remaining expressions,
 \item if the absolute value of each of the expressions 
 $< \epsilon$ $\approx$ ${\rm Machine Precision}/2$,
  then accept the solution as valid, else reject the solution
  (after showing it to the user).
\end{itemize}
{\it Mathematica} evaluates $\sqrt{a^2} \to a$ when $a$ is numeric, but 
does not evaluate $\sqrt{a^2} \to |a|$ when $a$ is symbolic.
Our simplification routines use $\sqrt{a^2} = a$ instead of $|a|$ when 
$a$ is symbolic. 
This has two consequences: (i) valid solutions may be missed, and 
(ii) solutions have a $1/2$ probability of evaluating to matching signs 
during the numeric test. 
The numeric test being inconclusive, we perform a symbolic test.
\vskip 1pt
\noindent
For the symbolic test of solutions:
\begin{itemize}
 \item after substituting the solution, expand and factor the left hand 
  side of (\ref{originalsystemPDEs}),
 \item apply simplification rules like $\sqrt{a^2} \to a,$
 $\sqrt{-a^2} \to i\,a,$
 $1-\sech^2\xi \to \tanh^2\xi,$ and
 $\sn^2(x;m) \to 1-\cn^2(x;m),$
 \item repeat the above simplifications until the expressions are static,
 \item if the final expressions are identically equal to zero,
 then accept the solution, else reject the solution and report
 the unresolved expressions to the user.
\end{itemize}
% 
%%%%%%%%%%%Examples of solitary wave solutions for ODEs and PDEs%%%%%%%%%%%
% 
\vspace{-0.25cm}
\section{Examples of solitary wave solutions for ODEs and PDEs}
\label{examplesPDEs}
The algorithms from Sections~\ref{tanhmethodPDEs}-\ref{cnmethodPDEs} 
were implemented in our {\em Mathematica} package 
\verb|PDESpecialSolutions.m|, 
which was used to solve the equations in this Section.
\vspace{-0.50cm}
\subsection{The Zakharov-Kuznetsov KdV-type equations}
\label{exampleZK}

The KdV-Zakharov-Kuznetsov (KdV-ZK) equation, 
\begin{equation}
\label{kdvzk}
u_t + \alpha u u_x + u_{xxx} + u_{xyy} + u_{xzz} = 0, 
\end{equation}
models ion-acoustic waves in magnetized multi-component plasmas including 
negative ions (see e.g.\ Das and Verheest, 1989).

With \verb|PDESpecialSolutions| (Tanh and Sech options) we found the solution
\begin{system}
\label{solskdvzk}
u(x,y,z,t) &=& 
\frac{8 c_1 (c_1^2 + c_2^2 + c_3^2) - c_4}{\alpha c_1} - 
\frac{12 (c_1^2 + c_2^2 + c_3^2)}{\alpha} \, {\tanh}^2\xi,  \\
&=& -\frac{4 c_1 (c_1^2 + c_2^2 + c_3^2) + c_4}{\alpha c_1} + 
\frac{12 (c_1^2 + c_2^2 + c_3^2)}{\alpha} \, {\sech}^2\xi, 
\end{system}
where $\xi = c_1 x + c_2 y + c_3 z + c_4 t + \Delta,$
with $c_1, c_2, c_3, c_4, \Delta $ and $\alpha$ arbitrary. 

For $c_2 = c_3 = 0$ and replacing $c_4$ by $c_2,$ one gets the solitary 
wave solution 
\begin{system}
\label{solskdv}
u(x,t) &=& \frac{8 c_1^3 - c_2}{\alpha c_1} - 
\frac{12 c_1^2}{\alpha} 
\, {\tanh}^2(c_1 x + c_2 t + \Delta),  \\
&=& -\frac{4 c_1^3 + c_2}{\alpha c_1} + 
\frac{12 c_1^2}{\alpha} \, {\sech}^2(c_1 x + c_2 t + \Delta ), 
\end{system}
of the ubiquitous KdV equation (\ref{orgkdv}).

The function \verb|PDESpecialSolutions| does not take boundary or initial 
conditions as input. 
One can {\rm a posteriori} impose conditions on solutions.
For example, requiring $\lim_{x \rightarrow \pm \infty}$ $u(x,t)$ $=0$ 
in (\ref{solskdv}) would fix $c_2 = - 4 c_1^3.$

For the modified KdV-ZK equation (Das and Verheest, 1989),
\begin{equation}
\label{mkdvzk}
u_t + \alpha u^2 u_x + u_{xxx} + u_{xyy} + u_{xzz} = 0, 
\end{equation}
using the Tanh and Sech options, \verb|PDESpecialSolutions| returns
\begin{equation}
\label{solsmkdvzktanh}
u(x,y,z,t) 
= \pm i \, \sqrt{{6 (c_1^2 + c_2^2 + c_3^2)}/{\alpha}} 
\, {\tanh}\,\xi, 
\end{equation}
with 
$\xi = c_1 x + c_2 y + c_3 z + 2 c_1 (c_1^2 + c_2^2 + c_3^2) t + \Delta,$ 
and 
\begin{equation}
\label{solsmkdvzksech}
u(x,y,z,t) = \pm \sqrt{{6 (c_1^2 + c_2^2 + c_3^2)}/{\alpha}}
\, {\sech}\,\xi, 
\end{equation}
with $\xi = c_1 x + c_2 y + c_3 z - c_1  (c_1^2 + c_2^2 + c_3^2) t + \Delta.$
For $c_2 = c_3 = 0,$ (\ref{solsmkdvzktanh}) and 
(\ref{solsmkdvzksech}) reduce to the well-known solitary wave solutions 
\begin{eqnarray}
\label{solmkdvtanh}
u(x,t) &=& \pm i c_1 \sqrt{{6}/{\alpha}} 
\, {\tanh}(c_1 x + 2 c_1^3 t + \Delta), \\
u(x,t) &=& \pm c_1 \sqrt{{6}/{\alpha}} 
\, {\sech}(c_1 x - c_1^3 t + \Delta)
\end{eqnarray}
($c_1, \Delta $ and $\alpha$ arbitrary real numbers) of the modified KdV 
(mKdV) equation (Ablowitz and Clarkson, 1991),    
\begin{equation}
\label{orgmkdv}
u_t + \alpha u^2 u_x + u_{xxx} = 0.
\end{equation}
For a three dimensional modified KdV (3D-mKdV) equation,
\begin{equation}
\label{3DmKdV}
u_t + 6 u^2 u_x + u_{xyz} = 0, 
\end{equation}
one obtains the solitary wave solution
\begin{equation}
\label{3DmKdVsechsolution}
u(x,y,z,t) = \pm \sqrt{c_2 c_3} 
\, {\sech}(c_1 \, x + c_2 \, y + c_3 \, z - c_1 c_2 c_3 \, t + \Delta),
\end{equation}
where $c_1, c_2, c_3$ and $\Delta$ are arbitrary.
\vspace{-0.50cm}
\subsection{The generalized Kuramoto-Sivashinsky equation}
\label{examplegenKdV}
Consider the generalized Kuramoto-Sivashinsky (KS) equation
(see e.g.\ Parkes and Duffy, 1996):
\begin{equation}
\label{ks}
u_t + u u_x + u_{xx} + \alpha u_{xxx} + u_{xxxx} = 0.
\end{equation}
Ignoring complex solutions, \verb|PDESpecialSolutions| (Tanh option)
automatically determines the special values of the real parameter 
$\alpha$ and the corresponding closed form solutions.
For $\alpha = 4,$
\begin{equation}
\label{kssol1}
u(x,t) = 9 \pm 2 c_2 \pm 15 \, {\tanh}\,\xi 
- 15 \, {\tanh}^2\xi \mp 15 {\tanh}^3\xi, 
\end{equation}
with $\xi = \mp \textstyle{\frac{1}{2}} x + c_2 t + \Delta.$
For $\alpha = \frac{12}{\sqrt{47}},$ 
\begin{equation}
\label{kssol2}
u(x,t) = \frac{45 \mp 4418 c_2}{47 \sqrt{47}} 
\pm \frac{45}{47 \sqrt{47}} 
\, {\tanh}\,\xi - \frac{45}{47 \sqrt{47}} 
\, {\tanh}^2\xi \pm \frac{15}{47 \sqrt{47}} 
\, {\tanh}^3\xi, 
\end{equation}
where $\xi = \pm \frac{1}{2 \sqrt{47}} x + c_2 t + \Delta.$
For $\alpha = \frac{16}{\sqrt{73}},$
\begin{equation}
\label{kssol3}
u(x,t) = \frac{2 (30 \mp 5329 c_2)}{73 \sqrt{73}} 
\pm \frac{75}{73 \sqrt{73}}
\, {\tanh}\,\xi - \frac{60}{73 \sqrt{73}} 
\, {\tanh}^2\xi \pm \frac{15}{73 \sqrt{73}} 
\, {\tanh}^3\xi, 
\end{equation}
where $\xi = \pm \frac{1}{2 \sqrt{73}} x + c_2 t + \Delta.$

The remaining solutions produced by \verb|PDESpecialSolutions| are either 
complex (not shown here) or can be obtained from the solutions above 
via the inversion symmetry of (\ref{ks}): 
$u \rightarrow -u, \, x \rightarrow -x, \, \alpha \rightarrow -\alpha.$

A separate run of the code after setting $\alpha = 0$ in (\ref{ks}) yields 
\begin{equation}
\label{kdsol4}
u(x,t) = - 2 \sqrt{\frac{19}{11}} c_2 
- \frac{135}{19}\sqrt{\frac{11}{19}} 
\, {\tanh}\,\xi 
+ \frac{165}{19}\sqrt{\frac{11}{19}} 
\, {\tanh}^3\xi, 
\end{equation} 
with $\xi = \textstyle{\frac{1}{2}}\sqrt{\frac{11}{19}} x + c_2 t + \Delta.$
In all solutions above $c_2$ is arbitrary.
\vspace{-0.50cm}
\subsection{Coupled KdV equations}
\label{examplecKdV}

In Section~\ref{sechmethodPDEs} we gave the sech-solutions for the 
Hirota-Satsuma system (\ref{orgckdv}).
Here we list the $\tanh, \cn$ and $\sn$ solutions for (\ref{orgckdv}) 
computed by \verb|PDESpecialSolutions| (Tanh, JacobiCN and JacobiSN options): 
\begin{system}
\label{tanhsolsckdv}
u(x,t) &=& \frac{2 c_1^3 - c_2}{3 c_1} - 2 c_1^2 
           \, {\tanh}^2(\xi), \\
v(x,t) &=& \pm \sqrt{[8 \alpha c_1^4 + 2 (1 + 2\alpha) c_1 c_2]/\beta} 
           \; {\tanh}(\xi), \\
u(x,t) &=& \frac{8 c_1^3 - c_2}{3 c_1} - 4 c_1^2 
           \, {\tanh}^2(\xi), \\
v(x,t) &=& 
 \pm \frac{8 \alpha c_1^3 - (1 + 2 \alpha) c_2}{c_1 \sqrt{6 \alpha \beta}} 
 \mp 2 c_1^2 \sqrt{6 \alpha/\beta} \; {\tanh}^2(\xi), 
\end{system}
\begin{system}
\label{snsolsckdv}
u(x,t) &=& \frac{(1 + m) c_1^3 - c_2}{3 c_1} - 2 m\,c_1^2 
           \, {\sn}^2(\xi; m ), \\
v(x,t) &=& \pm \sqrt{[4 \alpha m(1+m) c_1^4 
           + 2(1 + 2\alpha) m\,c_1 c_2]/\beta} 
           \; {\sn}(\xi, m), \\
u(x,t) &=& \frac{4 (1+m)c_1^3 - c_2}{3 c_1} - 4 m\,c_1^2 
           \, {\sn}^2(\xi; m), \\
v(x,t) &=& 
\pm \frac{4 \alpha (1+m) c_1^3 - (1+2 \alpha) c_2}{c_1 \sqrt{6 \alpha \beta}} 
 \mp 2 c_1^2 \sqrt{6 \alpha/\beta} \; {\sn}^2(\xi; m), \\
\end{system}
\begin{system}
\label{cnsolsckdv}
u(x,t)&=& \frac{(1 - 2 m) c_1^3 - c_2}{3 c_1} + 2 m\,c_1^2 
           \, {\cn}^2(\xi; m ), \\
v(x,t)&=& \pm \sqrt{[4 \alpha m(2m-1) c_1^4 -2(1 + 2\alpha) m\,c_1 c_2]/\beta} 
           \; {\cn}(\xi; m), \\
u(x,t)&=& \frac{4 (1-2m)c_1^3 - c_2}{3 c_1} + 4 m\,c_1^2 
           \, {\cn}^2(\xi; m), \\
v(x,t)&=& 
\pm \frac{4 \alpha (1-2m) c_1^3 - (1+2 \alpha) c_2}{c_1 \sqrt{6 \alpha \beta}} 
 \pm 2 c_1^2 \sqrt{6 \alpha/\beta} \; {\cn}^2(\xi; m), 
\end{system}
with $\xi = c_1 x + c_2 t + \Delta,$ and $c_1, c_2, \alpha, \beta, \Delta,$ 
and modulus $m$ arbitrary. 
These solutions correspond with those given in Fan and Hon (2002).

With the SechTanh option we obtained two dozen (real and complex) solutions. 
The real solutions coincide with the ones given above.
\vskip 3pt
\indent
Another coupled system of KdV-type equations was studied by Guha-Roy (1987)
\begin{system}
\label{guharoykdv}
&& u_t + \alpha v v_x + \beta u u_x + \gamma u_{xxx} = 0,  \\
&& v_t + \delta (u v)_x + \epsilon v v_{x} = 0,
\end{system}
where $\alpha$ through $\epsilon$ are real constants.
The package \verb|PDESpecialSolutions| (Sech option) computed:
\begin{system}
\label{guharoysol}
u(x,t) &=& 
-\frac{4\epsilon^2 \gamma c_1^3 + (4\alpha \delta + \epsilon^2)c_2 }{A c_1}
+ \frac{12\epsilon^2 \gamma c_1^2}{A} 
\, {\sech}^2(c_1 x + c_2 t + \Delta), \\
v(x,t) &=& 
\frac{2 \epsilon [4 \delta \gamma c_1^3 + (\delta - \beta )c_2 ] }{A c_1}
- \frac{24 \delta \epsilon \gamma c_1^2}{A} \, {\sech}^2(\xi),
\end{system}
where $\xi = c_1 x + c_2 t + \Delta, $
$ A =4\alpha\delta^2 + \beta\epsilon^2,$ with $c_1, c_2, \Delta$
and $\alpha$ through $\epsilon$ arbitrary.
For $\epsilon = 0,$ (\ref{guharoykdv}) reduces to Kawamoto's system;
for $\epsilon = 0,\delta = -2$ to Ito's system. 
Neither of these systems has polynomial solutions in 
$\sech$ or $\tanh.$
\vspace{-0.50cm}
\subsection{The Fisher and FitzHugh-Nagumo equations}
\label{examplefisher}

For the Fisher equation (Malfliet, 1992), 
\begin{equation}
\label{fisher}
u_t - u_{xx} - u (1 - u) = 0, 
\end{equation}
with \verb|PDESpecialSolutions| (Tanh option) we found the (real) solution
\begin{equation}
\label{fishersol}
u(x,t) = \textstyle{\frac{1}{4}} \pm \textstyle{\frac{1}{2}} 
\, {\tanh}\,\xi 
+ \textstyle{\frac{1}{4}} \, {\tanh}^2\xi,  
\end{equation}  
with $\xi = \pm \frac{1}{2 \sqrt{6}}x \pm \frac{5}{12} t + \Delta.$ 
In addition, there are 4 complex solutions.

Obviously, \verb|PDESpecialSolutions| handles ODEs also. 
For example, we can put the FitzHugh-Nagumo (FHN) equation (Hereman, 1990),
\begin{equation}
\label{fhnpde}
u_t - u_{xx} + u (1 - u) (\alpha - u) = 0,
\end{equation}
where $-1 \leq \alpha < 1,$ into a travelling frame, 
\begin{equation}
\label{fhnode}
\beta v_z + \sqrt{2} v_{zz} - 
\sqrt{2} v (1 - \sqrt{2} v) (\alpha - \sqrt{2} v) = 0, 
\end{equation}
with $v(z) = v(x-\frac{\beta}{\sqrt{2}} t) = \sqrt{2}\, u(x,t).$ 
Ignoring the inversion symmetry 
$z \rightarrow -z, \, \beta \rightarrow -\beta$ of (\ref{fhnode}), 
we find with \verb|PDESpecialSolutions| (Tanh option)
\begin{equation}
\label{solutionfhnode1}
v(z)=\frac{1}{2 \sqrt{2}} \left[ \beta + (\beta-2) 
\, {\tanh}[\frac{\sqrt{2}}{4}(2-\beta) z + \Delta ] \right], 
\end{equation}
if $\alpha = \beta-1;$
\begin{equation}
\label{solutionfhnode2}
v(z)=\frac{(\beta + 2)}{2\sqrt{2}} 
\left[ 1 - {\tanh}[\frac{\sqrt{2}}{4}(\beta +2) z + \Delta ] \right], 
\end{equation}
if $\alpha = \beta+2;$ and
\begin{equation}
\label{solutionfhnode3}
v(z)=\frac{1}{2 \sqrt{2}} 
\left[ 1 + {\tanh}[\frac{\sqrt{2}}{4} z + \Delta ] \right], 
\end{equation}
if $\alpha = \textstyle{\frac{1}{2}}(\beta+1).$
In these solutions (see e.g.\ Hereman, 1990) $\beta$ and $\Delta$ are 
arbitrary.
\vspace{-0.50cm}
\subsection{A degenerate Hamiltonian system}
\label{completelyintegrable}

Gao and Tian (2001) considered the following degenerate Hamiltonian system, 
\begin{system}
\label{orghamiltonianeqs}
&& u_t - u_x - 2 v = 0, \\
&& v_t - 2 \epsilon\, u v = 0,  \quad \epsilon = \pm 1, 
\end{system}
which was shown to be completely integrable by admitting infinitely many 
conserved densities.
Our code does not find sech-solutions. 
With the SechTanh option, \verb|PDESpecialSolutions| returns the solutions:
\begin{system}
\label{hamiltonianeqssolution1}
u(x,t) &=& - \epsilon c_2 \, {\tanh}\,\xi,  \\
v(x,t) &=& \textstyle{\frac{1}{2}} \epsilon\, c_2 (c_1 - c_2) \, {\sech}^2\xi, 
\end{system}
which could have been obtained with the tanh-method in 
Section~\ref{tanhmethodPDEs}; and
\begin{system}
\label{hamiltonianeqssolution2}
u(x,t) &=& \textstyle{\frac{1}{2}} i c_2 \epsilon \,
({\sech}\,\xi + i \, {\tanh}\,\xi ),  \\
v(x,t) &=& \textstyle{\frac{1}{4}} c_2 (c_1 - c_2) \epsilon\; {\sech}\,\xi 
\, ({\sech}\,\xi + i \, {\tanh}\,\xi ), 
\end{system}
plus their two complex conjugates. 
There are no constraints on $c_1, c_2,$ and $\epsilon,$
and $\xi = c_1 x + c_2 t + \Delta.$
The above solutions were reported in Gao and Tian (2001).
\vspace{-0.50cm}
\subsection{The combined KdV-mKdV equation}
\label{kdvmkdvexample}

The combined KdV-mKdV equation (see Gao and Tian, 2001)  
\begin{equation}
\label{kdvmkdvorg}
u_t + 6 \alpha u u_x + 6 \beta u^2 u_x + \gamma u_{xxx} = 0, 
\end{equation}
describes a variety of wave phenomena in plasma, solid state, and 
quantum physics. 
We chose this example to show that ODEs of type (\ref{legendretypesech}), 
which are {\rm free} of $\sqrt{1-S^2},$ can admit mixed tanh-sech solutions. 

First, \verb|PDESpecialSolutions| with the Tanh option, produces
\begin{equation}
\label{kdvmkdvsol1}
u(x,t) = -\frac{\alpha}{2 \beta} \pm i \sqrt{\frac{\gamma}{\beta}} c_1
\, {\tanh}(c_1 x + \frac{c_1}{2 \beta} (3 \alpha^2 + 4 \beta \gamma c_1^2) t 
+ \Delta).
\end{equation}
Next, with the Sech option, \verb|PDESpecialSolutions| computes
\begin{equation}
\label{kdvmkdvsol2}
u(x,t) = -\frac{\alpha}{2 \beta} \pm \sqrt{\frac{\gamma}{\beta}} c_1
\, {\sech}[c_1 x + \frac{c_1}{2 \beta} (3 \alpha^2 - 2 \beta \gamma c_1^2) t 
+ \Delta ].
\end{equation}
Third, with the SechTanh option, \verb|PDESpecialSolutions| finds
\begin{equation}
\label{kdvmkdvsol3}
u(x,t) = -\frac{\alpha}{2 \beta} + 
\textstyle{\frac{1}{2}} \sqrt{\frac{\gamma}{\beta}} c_1 \, ({\sech}\,\xi 
\pm i\, {\tanh}\,\xi), 
\end{equation}
and 
\begin{equation}
\label{kdvmkdvsol4}
u(x,t) = -\frac{\alpha}{2 \beta} -
\textstyle{\frac{1}{2}} \sqrt{\frac{\gamma}{\beta}} c_1 \, ({\sech}\,\xi 
\mp i\,{\tanh}\,\xi),
\end{equation}
where $\xi = c_1 x + \textstyle{\frac{1}{2}} \frac{c_1}{\beta} 
(3 \alpha^2 +\beta \gamma c_1^2)t +\Delta.$
In all solutions $c_1, \Delta, \alpha, \beta $ and $\gamma$ are arbitrary. 
The solutions were reported in Gao and Tian (2001), although there 
were minor misprints.  
\vspace{-0.50cm}
\subsection{The Duffing Equation}
\label{exampleduffing}

Duffing's equation (Lawden, 1989),  
\begin{equation}
\label{duffing}
u^{\prime\prime} + u + \alpha u^3 = 0, 
\end{equation}
models a nonlinear spring problem. 
Its $\cn$ and $\sn$ solutions
\begin{system}
\label{duffingcnandsnsolution}
u(x) &=& \pm \sqrt{\frac{2m}{(1-2 m) \alpha}} \, 
{\cn}(\frac{\epsilon x}{\sqrt{1-2m}} + \Delta ; m), \; \epsilon = \pm 1, \\
u(x) &=& \pm \sqrt{\frac{-2m}{(1+m)\alpha}} \,
{\sn}(\frac{\epsilon x}{\sqrt{1+m}} + \Delta ; m), \; \epsilon = \pm 1,
\end{system} 
are computed by \verb|PDESpecialSolutions| with the JacobiCN and JacobiSN 
options. 
There are four sign combinations in (\ref{duffingcnandsnsolution}).
Since $0 \leq m \leq 1,$ the $\cn$ solution is real when $\alpha > 0$ and 
$m < \textstyle{\frac{1}{2}}.$ 
The $\sn$ solution is real for $\alpha < 0.$ 
Such conditions are not automatically generated.
During simplifications the code assumes $\alpha > 0$
(see Section~\ref{analyzeandsolve} for details).

Initial conditions fix the modulus in (\ref{duffingcnandsnsolution}).
For example, $u(0)=a$ and $\dot{u}(0)=0$ lead to
$u(x)=a\,{\cn}(\sqrt{1+\alpha a^2}x; {(\alpha a^2)}/{(2+2\alpha a^2)}).$
\vspace{-0.50cm}
\subsection{A class of fifth-order PDEs with three parameters}
\label{examplefifthKdV}

To illustrate the limitations of \verb|PDESpecialSolutions| consider 
the family of fifth-order KdV equations (G\"okta\c{s} and Hereman, 1997), 
\begin{equation} 
\label{KdV5par}
u_t + \alpha u^2 u_{x} + \beta u_x u_{xx} + \gamma u u_{xxx} + u_{xxxxx} = 0,
\end{equation}
where $\alpha, \beta,$ and $\gamma$ are nonzero parameters.

An investigation of the scaling properties of (\ref{KdV5par}) reveals that 
only the ratios $\frac{\alpha}{\gamma^2}$ and $\frac{\beta}{\gamma}$ are 
important, but let us proceed with (\ref{KdV5par}). 
\vskip 3pt
\noindent
{\bf Special cases}
\vskip 1.5pt
\noindent
Several special cases of (\ref{KdV5par}) are well known (for references 
see G\"okta\c{s} and Hereman, 1997).
Indeed, for $\alpha = 30, \beta = 20, \gamma = 10,$ 
Eq.\ (\ref{KdV5par}) reduces to 
\begin{equation}
\label{lax}
u_t + 30 u^2 u_{x} + 20 u_x u_{xx} + 10 u u_{xxx} + u_{xxxxx} = 0,
\end{equation}
which belongs to the completely integrable hierarchy of higher-order 
KdV equations constructed by Lax. 
Equation (\ref{lax}) has two tanh-solutions: 
\begin{equation}
\label{laxsol1}
u(x,t) = 4 c_1^2 - 6 c_1^2 \, {\tanh}^2(c_1 x - 56 c_1^5 t + \Delta),
\end{equation}
and 
\begin{equation}
\label{laxsol2}
u(x,t) = a_{10} - 2 c_1^2 
\, {\tanh}^2[ c_1 x - 2 (15 a_{10}^2 c_1 -40 a_{10} c_1^3 +28 c_1^5)t +
\Delta ], 
\end{equation}
where $a_{10}, c_1, \Delta$ are arbitrary. 

For $\alpha = \beta = \gamma = 5,$ one gets the equation, 
\begin{equation}
\label{sk}
u_t + 5 u^2 u_{x} + 5 u_x u_{xx} + 5 u u_{xxx} + u_{xxxxx} = 0,
\end{equation}
due to Sawada and Kotera (SK) and  Dodd and Gibbon, 
which has tanh-solutions: 
\begin{equation}
u(x,t) = 8 c_1^2 - 12 c_1^2 \, {\tanh}^2(c_1 x - 16 c_1^5 t + \Delta),
\end{equation}
and 
\begin{equation}
u(x,t) = a_{10} - 6 c_1^2 \, {\tanh}^2[ c_1 x - (5 a_{10}^2 c_1 - 
40 a_{10} c_1^3 + 76 c_1^5) t + \Delta],
\end{equation}
where $a_{10}, c_1, \Delta$ are arbitrary. 

The KK equation due to Kaup and Kupershmidt, 
\begin{equation}
\label{kk}
u_t + 20 u^2 u_{x} + 25 u_x u_{xx} + 10 u u_{xxx} + u_{xxxxx} = 0,
\end{equation}
corresponding to $\alpha = 20, \beta = 25, \gamma = 10,$ 
and again admits two tanh-solutions: 
\begin{equation}
\label{kksol1}
u(x,t) = c_1^2 - 
\frac{3}{2} c_1^2 \, {\tanh}^2(c_1 x - c_1^5 t + \Delta),
\end{equation}
and 
\begin{equation}
\label{kksol2}
u(x,t) = 8 c_1^2 - 12 c_1^2 \, {\tanh}^2(c_1 x - 176 c_1^5 t + \Delta), 
\end{equation}
with $c_1, \Delta$ arbitrary, but no additional arbitrary coefficients. 

The equation 
\begin{equation}
\label{ito}
u_t + 2 u^2 u_{x} + 6 u_x u_{xx} + 3 u u_{xxx} + u_{xxxxx} = 0,
\end{equation}
for $\alpha = 2, \beta = 6, \gamma = 3,$ 
was studied by Ito. It has one $\tanh$ solution: 
\begin{equation}
\label{itosol1}
u(x,t) = 20 c_1^2 - 30 c_1^2 \, {\tanh}^2(c_1 x - 96 c_1^5 t + \Delta), 
\end{equation}
again with $c_1$ and $\Delta$ arbitrary.
\verb|PDESpecialSolutions| (Tanh option) produces all these solutions. 
\vskip 3pt
\noindent
{\bf General case}
\vskip 1.5pt
\noindent
Eq.\ (\ref{KdV5par}) is hard to analyze by hand or with the computer. 
After a considerably amount of time, \verb|PDESpecialSolutions|
(Tanh option) produced the solutions given below (but not in as nice a form). 
Our write-up of the solutions is the result of additional interactive work 
with {\em Mathematica}.

The coefficients $a_{10}, a_{11},$ and $a_{12}$ in 
\begin{equation}
\label{generalsol1}
u(x,t) = a_{10} + a_{11} \, {\tanh}(\xi) +
a_{12} \, {\tanh}^2(\xi), 
\end{equation}  
with $\xi = c_1 x + c_2 + \Delta,$ must satisfy the following nonlinear 
algebraic system with parameters $c_1, c_2, \alpha, \beta, $ and $\gamma:$
\begin{system}
\label{5kdvparsys}
\alpha a_{12}^2 + 6 \beta a_{12} c_1^2 
+ 12 \gamma a_{12} c_1^2 + 360 c_1^4 &=& 0,  \\
a_{11} \, ( \alpha a_{12}^2 + 2 \beta a_{12} c_1^2 
+ 6 \gamma a_{12} c_1^2 + 24 c_1^4 ) &=& 0,  \\
a_{11} \, ( \alpha a_{10}^2 c_1 - 2 \gamma a_{10} c_1^3 
+ 2 \beta a_{12} c_1^3 + 16 c_1^5 + c_2 ) &=& 0,  \\
a_{11} \, ( \alpha a_{11}^2 + 6 \alpha a_{10} a_{12} 
+ 6 \gamma a_{10} c_1^2 - 12 \beta a_{12} c_1^2 
- 18 \gamma a_{12} c_1^2 - 120 c_1^4 ) &=& 0,  \\ 
2 \alpha a_{11}^2 a_{12} + 2 \alpha a_{10} a_{12}^2 
+ \beta a_{11}^2 c_1^2 + 3 \gamma a_{11}^2 c_1^2 
+ 12 \gamma a_{10} a_{12} c_1^2 && \\
\qquad \quad- 8 \beta a_{12}^2 c_1^2 - 8 \gamma a_{12}^2 c_1^2 
- 480 a_{12} c_1^4 &=& 0,  \\
\alpha a_{10} a_{11}^2 c_1 + \alpha a_{10}^2 a_{12} c_1 
- \beta a_{11}^2 c_1^3 - \gamma a_{11}^2 c_1^3 
- 8 \gamma a_{10} a_{12} c_1^3 + 2 \beta a_{12}^2 c_1^3  && \\
\qquad \quad + 136 a_{12} c_1^5 + a_{12} c_2 &=& 0.
\end{system}
Assuming nonzero $a_{12}, c_1, c_2, \alpha, \beta,$ and $\gamma,$ 
two cases must be distinguished:
\vskip 3pt
\noindent
{\bf Case 1:} $a_{11} = 0.\;$
In turn, this case splits into two sub-cases:
\vskip 2pt
\noindent
{\bf Case 1a:}
\vskip 1pt
\noindent
\begin{equation}
\label{5kdvcase1a}
a_{11} = 0, \; a_{12} = - \frac{3}{2} a_{10}, \;
c_2 = c_1^3 ( 24 c_1^2 - \beta a_{10} ), 
\end{equation}
where $a_{10}$ must be one of the roots of 
\begin{equation}
\label{case1aquad}
\alpha a_{10}^2 - 4 \beta a_{10} c_1^2 - 8 \gamma a_{10} c_1^2 
+ 160 c_1^4 = 0.
\end{equation} 
\vskip 2pt
\noindent
{\bf Case 1b:}
\vskip 1pt
\noindent
\begin{equation}
\label{5kdvcase1b}
a_{11} \!=\! 0,\; 
a_{12} \!=\! -\frac{6 \gamma}{\alpha} c_1^2,\;
c_2 \!=\!-\frac{1}{\alpha} 
(\alpha^2 a_{10}^2 c_1 \!-\! 8 \alpha \gamma a_{10} c_1^3 
\!+\! 16 \alpha c_1^5 \!+\! 12 \gamma^2 c_1^5 ),
\end{equation}
provided that
\begin{equation}
\label{condcase1a}
\beta = \frac{1}{\gamma}{(10 \alpha - \gamma^2)}. 
\end{equation}
\vskip 3pt
\noindent
{\bf Case 2:} $a_{11} \ne 0.\;$
Then 
\begin{equation}
\label{condcase2first}
\alpha = \frac{1}{392} (8 \beta^2 + 38 \beta \gamma + 39 \gamma^2), \;\;
a_{12} = -\frac{168}{2 \beta + 3 \gamma } c_1^2,
\end{equation}
provided $\beta$ is one of the roots of 
\begin{equation}
\label{condcase2third}
(104 \beta^2 + 886 \beta \gamma + 1487 \gamma^2 )
(520 \beta^3 + 2158 \beta^2 \gamma - 1103 \beta \gamma^2 - 8871 \gamma^3 )=0.
\end{equation}
Thus, case 2 also splits into two sub-cases:
\vskip 3pt
\noindent
{\bf Case 2a:}
If $\beta^2 = -\frac{1}{104} (886 \beta \gamma + 1487 \gamma^2),$ then
\begin{system}
\label{case2aquad}
&& \alpha = - \frac{1}{26} (2 \beta + 5 \gamma) \gamma,  \;\;
a_{10} = -\frac{52 (4378 \beta + 9983 \gamma}
{7 \gamma (958 \beta + 2213 \gamma )} c_1^2, \;\;
a_{11} = \pm \frac{336}{2 \beta + 3 \gamma} c_1^2, \\
&& a_{12} = - \frac{168}{2 \beta + 3 \gamma} c_1^2,  \;\;
c_2 = -\frac{364 (1634 \beta + 3851 \gamma)}{2946 \beta + 6715 \gamma} c_1^5.
\end{system}
where $\beta$ is any root of 
$104 \beta^2 + 886 \beta \gamma + 1487 \gamma^2 =0.$
\vskip 3pt
\noindent
{\bf Case 2b:}
If 
$\beta^3 = 
\frac{1}{520} (1103 \beta \gamma^2 + 8871 \gamma^3 - 2158 \beta^2 \gamma ),$ 
then
\begin{system}
\label{case2cubic}
&& \alpha = \frac{1}{392} 
(8 \beta^2 + 38 \beta \gamma + 39 \gamma^2),  \;\;
a_{10} = 
\frac{28 (1066 \beta^2 + 5529 \beta \gamma + 6483 \gamma^2)}
{(2 \beta + 3 \gamma)(6 \beta + 23 \gamma)(26 \beta + 81 \gamma)} c_1^2,  \\
&& a_{11}^2 = \frac{28224 (26 \beta - 17 \gamma) 
(4 \beta - \gamma)} {(2 \beta + 3 \gamma)^2 (6 \beta + 23 \gamma)
(26 \beta + 81 \gamma)} c_1^4, \;\;
a_{12} = - \frac{168}{2 \beta + 3 \gamma} c_1^2,  \\
&& c_2 = -\frac{8 (188900114 \beta^2 + 1161063881 \beta \gamma + 
   1792261977 \gamma^2)}{105176786 \beta^2 + 
   632954969 \beta \gamma + 959833473 \gamma^2} c_1^5,
\end{system}
where $\beta$ is any root of 
$520 \beta^3 +2158 \beta^2 \gamma -1103 \beta \gamma^2 -8871 \gamma^3 = 0.$
% 
% 
%%%%%%%%%%%%%Other Algorithms and Related Software%%%%%%%%%%%%%%%
% 
\vspace{-0.25cm}
\section{Other Algorithms and Related Software}
\label{relatedalgossoftware}
% 
%%%%%%%%%%%%%Other perspectives and potential generalizations%%%%%%%%%%%%%%%
% 
% \vspace{-0.50cm}
\subsection{Other perspectives and potential generalizations}
\label{generalizations}

The algorithms presented in this article can be extended in several ways.
For instance, one could modify the chain rule in step T1 (S1, ST1, or CN1) 
to compute other {\em types} of solutions or consider more complicated 
polynomials than those used in step T2 (S2, ST2, or CN2).
Both options could be used together.

With respect to the first option, it suffices to know the underlying 
first-order differential equation of the desired {\rm fundamental} 
function in the polynomial solution. 
Table~\ref{tbl:functionsODEs} summarizes some of the more obvious choices. 
\vskip 0.0001pt
\noindent
\begin{table}
\vspace{-0.20cm}
\centering
\begin{tabular}{|l|l|l|l|} 
\hline 
Function &$\!\!$Symbol$\!\!$& ODE $\;\;(y'\!=\!\frac{dy}{d\xi})$ 
& Chain Rule 
\rule[-8pt]{0.0in}{20pt} \\
\hline 
${\tanh}(\xi)$ &\mbox{\scriptsize{T}} & $y^{\prime}\!=\!1-y^2$ 
& $\frac{\partial \bullet}{\partial x_j} 
\!=\!c_j(1-\mbox{\scriptsize{T}}^2)\frac{d\bullet}{d\mbox{\scriptsize{T}}}$
\rule[-8pt]{0.0in}{20pt} \\
\hline 
$ {\sech}(\xi)$ & \mbox{\scriptsize{S}} & $y^{\prime}\!=\!-y \sqrt{1-y^2}$ 
&$\frac{\partial\bullet}{\partial x_j}
\!=\!-c_j \mbox{\scriptsize{S}}
\sqrt{1-\mbox{\scriptsize{S}}^2}\frac{d\bullet}{d\mbox{\scriptsize{S}}}$
\rule[-8pt]{0.0in}{20pt} \\
\hline 
${\tan}(\xi)$ & \mbox{\scriptsize{TAN}} & $y^{\prime}\!=\!1 + y^2 $ 
& $\frac{\partial \bullet}{\partial x_j} 
\!=\! c_j (1+ \mbox{\scriptsize{TAN}}^2) 
\frac{d \bullet}{d \mbox{\scriptsize{TAN}}}$ 
\rule[-8pt]{0.0in}{20pt} \\
\hline 
$\exp(\xi)$ & \mbox{\scriptsize{E}} & $y^{\prime}\!=\!y$ 
& $\frac{\partial \bullet}{\partial x_j} 
\!=\!c_j \mbox{\scriptsize{E}} 
\frac{d \bullet}{d \mbox{\scriptsize{E}}}$
\rule[-8pt]{0.0in}{20pt} \\
\hline 
$\cn(\xi;m)$ & \mbox{\scriptsize{CN}} & $y^{\prime} 
\!=\!-\!\sqrt{(1\!-\!y^2)(1\!-\!m\!+\!m\,y^2)}$ 
& $\frac{\partial \bullet}{\partial x_j}\!=\! 
\!-\!c_j\sqrt{(1\!-\!\mbox{\scriptsize{CN}}^2)
(1\!-\!m\!+\!m\,\mbox{\scriptsize{CN}}^2)} 
\frac{d\bullet}{d\mbox{\scriptsize{CN}}}\!$
\rule[-8pt]{0.0in}{20pt} \\
\hline 
$\sn(\xi;m)$ & \mbox{\scriptsize{SN}} & $y^{\prime} 
\!=\!\sqrt{(1 - y^2)(1 - m\,y^2)}$ 
& $\frac{\partial \bullet}{\partial x_j} 
\!=\!c_j\sqrt{(1-\mbox{\scriptsize{SN}}^2)
(1-m \mbox{\scriptsize{SN}}^2)}
\frac{d \bullet}{d \mbox{\scriptsize{SN}}}$ 
\rule[-8pt]{0.0in}{20pt} \\
\hline 
$\!{\cal P}(\xi;g_2,g_3)\!$ & \mbox{\scriptsize{P}} & $y^{\prime} 
\!=\!\pm \sqrt{4 y^3 -g_2 y -g_3}$ 
& $\frac{\partial \bullet}{\partial x_j}  
\!=\!\pm c_j \sqrt{4 y^3 \!-\!g_2 y \!-\!g_3} 
\frac{d \bullet}{d \mbox{\scriptsize{P}}}$ 
\rule[-8pt]{0.0in}{20pt} \\
\hline 
\end{tabular}
\caption{
Functions with corresponding ODEs and chain rules. 
${\cal P}(x;g_2,g_3)$ is the Weierstrass function with invariants 
$g_2$ and $g_3.$
}
\label{tbl:functionsODEs}
\end{table}
\vskip 0.0001pt
\noindent
Several researchers, including Fan (2002abc) and Gao and Tian (2001),
seek solutions of the form
\begin{equation}
\label{seriesU}
u_i(x,t) = U_i(\xi) = \sum_{j=1}^{M_i} a_{ij} w(\xi)^j, \quad
\xi = c_1 x  + c_2 t + \Delta,
\end{equation} 
where $w(\xi)$ is constrained by a Riccati equation,
\begin{equation}
\label{riccatitanhtan}
w^{\prime}(\xi) = b + \epsilon w^2(\xi), 
\,\;\epsilon = \pm 1, \,\;b\; {\rm real \, constant.}
\end{equation}
Ignoring rational solutions, (\ref{riccatitanhtan}) has the following 
solutions
\begin{system}
w(\xi) &=& a\,\tanh (a \xi + c), \quad {\rm if}\; \epsilon=-1, \; b = a^2, \\
w(\xi) &=& a\,\coth (a \xi + c), \quad {\rm if}\; \epsilon=-1, \; b = a^2, \\
w(\xi) &=& a\,\tan (a \xi + c),\;\, 
\quad {\rm if}\;\; \epsilon= 1, \; b = a^2, \\
w(\xi) &=& a\,\cot (a \xi + c), \;\, 
\quad {\rm if}\;\; \epsilon=-1, \; b = -a^2. 
\end{system}
So, (\ref{seriesU}) is polynomial in ${\tanh}\,\xi,$ $ {\tan}\,\xi, $ 
$ {\coth}\,\xi,$ or ${\cot}\,\xi.$
The integration constant $c$ gets absorbed in $\Delta,$ and the constant 
$a$ (or $b)$ is an extra parameter in the nonlinear algebraic system for 
the $a_{ij}.$ 
For single PDEs, Yao and Li (2002ab) consider solutions of the form
\begin{equation}
\label{sechtanhcombination}
u(x,t) = U(\xi) = \sum_{j=0}^{M} a_{j} w(\xi)^j + 
\sum_{j=0}^{M} b_{j} z(\xi) w(\xi)^{j-1},
\end{equation}
where $w(\xi)$ and $z(\xi)$ satisfy the Riccati equations
\begin{equation}
\label{riccatisystem}
w^{\prime}(\xi) = - w(\xi) z(\xi), \quad
z^{\prime}(\xi) = 1 - z^2(\xi).
\end{equation}
Since $w(\xi) = \sech(\xi), z(\xi) = \tanh(\xi)$ this approach is 
similar to the sech-tanh method in Section~\ref{sechtanhmethodPDEs}.

Generalizing further, Fan (2002b, 2003abc), Fan and Hon (2002, 2003a),
and Hon and Fan (2004b) take
\begin{equation}
\label{generakriccati}
y^{\prime}(\xi) = \sqrt{b_0 + b_1 y + b_2 y^2 + b_3 y^3 + b_4 y^4}, 
  \;\;\; b_i\;\; {\rm constant,}
\end{equation}
which covers the functions $\sech, \sec, \tanh, \tan, \cn, \sn,$ and 
${\cal P}.$
The parameters $b_i$ are added to the nonlinear algebraic system, 
which makes such systems hard to solve without human intervention. 
Most often, such complicated nonlinear algebraic systems are solved 
interactively with the aid of {\em Mathematica} or {\em Maple}. 
To avoid unmanageable systems, $M_i ( \leq 2)$ is often fixed 
in (\ref{seriesU}).
Chen and Zhang (2003ab), Fan and Dai (2003), and Sirendaoreji (2003, 2004) 
use variants of (\ref{generakriccati}) to compute polynomial and 
rational solutions in terms of $\tanh, \sech, \tan,$ Jacobi's 
elliptic functions, etc.

Zheng {\it et al.\/} (2002) introduce a clever method to compute mixed 
tanh-sech solutions for the combined KdV-Burgers equations. 
They seek formal solutions,  
\begin{equation}
\label{sincoscombination}
u(x,t) = U(\xi) = a_0 + \sum_{j=1}^{M} b_{j} \sin^{j} w(\xi) 
+ \sum_{j=1}^{M} a_{j} \cos w(\xi) \sin w(\xi)^{j-1}, 
\end{equation}
subject to $\frac{dw}{d\xi} = \sin w(\xi)$ which, upon integration,
gives $\sin w(\xi) = \sech(\xi)$ and $\cos w(\xi) = \pm \tanh(\xi).$
Alternatively, one can use $\frac{dw}{d\xi} = \cos w(\xi),$
which leads to $\cos w(\xi) = -\sech(\xi)$ and $\sin w(\xi) = \pm \tanh(\xi).$

Liu and Li (2002a) seek solutions of the forms
\begin{system}
\label{cnsncombination}
U(\xi) &=& \sum_{j=0}^{M} a_{j} \sn(\xi)^j,\quad 
U(\xi) = \sum_{j=0}^{M} a_{j} \sn(\xi)^j + 
\sum_{j=0}^{M} b_{j} \cn(\xi) \sn(\xi)^{j-1},  \\
U(\xi) &=& \sum_{j=0}^{M} a_{j} \sn(\xi)^j
+ \sum_{j=0}^{M} A_{j} \cn(\xi)\sn(\xi)^{j-1} 
+ \sum_{j=0}^{M} b_{j} \dn(\xi) \sn(\xi)^{j-1} \\
&& +\sum_{j=0}^{M} B_{j} \cn(\xi) \dn(\xi) \sn(\xi)^{j-2}, 
\end{system}
which generalize the Jacobi elliptic function method in 
Section~\ref{sechtanhmethodPDEs}. 

With respect to the second option, Gao and Tian (2001) consider
\begin{equation}
\label{sechtanhsolutions}
u_i(x,t) = \sum_{j=0}^{M_i} a_{ij}(x,t) \, {\tanh}^j\Psi(x,t) + 
\sum_{j=0}^{N_i} b_{ij}(x,t) \, {\sech}\Psi(x,t) \, {\tanh}^j\Psi(x,t),
\end{equation}
where $\Psi(x,t)$ is not necessarily linear in $x$ and/or $t.$
Of course, (\ref{sechtanhsolutions}) arises from recasting the terms
in (\ref{generalsolutionsechtanhPDEs}) in a slightly different way
than (\ref{polynomialsolutionsechtanhPDEs}). 
Restricted to travelling waves, 
$\Psi(x,t) = c_1 x + c_2 t + \Delta,$ both forms are equivalent. 

Our algorithms could be generalized in many ways.
With considerable effort, solutions involving complex exponentials 
multiplied by $\tanh$ or $\sech$ functions could be attempted. 
A solution to the nonlinear Schr\"odinger equation is of this form. 
Fan and Hon (2003b), Hon and Fan (2004a) and Fan (2003bc) 
give examples of complex as well as transcendental equations solved 
with the tanh-method.
% 
%%%%%%%%%%%%%%%%%%%%%%%%%%%%%%%%Review Software%%%%%%%%%%%%%%%%%%%%%%%%%%%%%%
% 
% \vspace{-0.25cm}
\subsection{Review of Symbolic Algorithms and Software}
\noindent
\label{reviewsoftware}
There is a variety of methods to find solitary wave solutions and
soliton solutions of special nonlinear PDEs. 
See e.g.\ Hereman and Takaoka (1990), Est\'evez and Gordoa (1995, 1998), 
and Helal (2002).
Some of these methods are straightforward to implement in computer 
algebra systems (CAS). 

The most comprehensive methods of finding exact solutions for ODEs and PDEs 
are based on similarity reductions via Lie point symmetry methods.
These methods are hard to fully automate
(for publications and software see 
e.g.\ Cantwell, 2002; Hereman, 1996; Hydon, 2000).
Most CAS have tools to solve a subset of linear and nonlinear PDEs. 
For example, {\em Mathematica's} \verb|DSolve| can find general solutions for 
linear and weakly nonlinear PDEs.
Available within {\em MuPAD}, the code \verb|pdesolve| uses the method 
of characteristics to solve quasi-linear first order PDEs.
{\em Maple} offers the packages {\rm ODEtools} (for solving ODEs using 
classification, integrating factor and symmetry methods) and 
{\rm PDEtools}, which contains the function {\em pdesolve} to find exact 
solutions of some classes of PDEs.
For information consult Cheb-Terrab (1995, 2001ab). 

The methods presented in this paper are different from these efforts.
Our algorithms and software only compute specific solutions of nonlinear PDEs
which model travelling waves in terms of the 
$\tanh, \sech, \sn$ and $\cn$ functions.
Our code can handle systems of ODEs and PDEs with undetermined parameters.

To our knowledge, only four software packages are similar to ours. 
The first package is {\rm ATFM} by Parkes and Duffy (1996), 
who automated to some degree the tanh-method using {\em Mathematica}. 
In contrast to {\rm ATFM}, our software performs the computations 
automatically from start to finish without human intervention. 
In our code, the number of independent variables $x_i$ is not limited to one 
space variable $x$ and time $t;$ our code handles any number of 
dependent variables.

The second package is {\rm RATH} by Li and Liu (2002), which automates 
the tanh-method.
In contrast to our code, {\rm RATH} only works for single PDEs.
Extensions to cover systems of PDEs and $\sech$ solutions are 
under development. 
Surpassing our code, {\rm RATH} can solve PDEs with an unspecified degree of 
nonlinearity and deal with negative and fractional exponents.

Table~\ref{tbl:comparison} compares the performance of 
\verb|PDESpecialSolutions.m| and {\rm RATH}.
The solution times are comparable, yet occasionally there is a mismatch
in the number of solutions computed. 
This is due in part to the representation of solutions.
Occasionally special solutions are generated 
although--after inspection by hand--they are included in more 
general solutions.
\vskip 0.0001pt
\noindent
\begin{table}
\centering
\begin{tabular}{r|c|c|c|c|l}
 & \multicolumn{2}{|c}{PDESpecialSolutions.m} & 
\multicolumn{2}{|c|}{RATH} & % \multicolumn{1}{|l}{Ref. \/} 
\\ \cline{2-5} %\hline
Name of Equation & CPU time & \# sols.\/ & CPU time & \# sols.\/ & Ref.\/ 
\\ \hline \hline
KdV-Burgers & 0.125s & 2 & 0.328s & 1 & (2.3) \\ \hline
KdV-Burgers-Kuramoto & 0.390s & 8 & 25.641s & 7 & (4.1) \\ \hline
7th-order dispersive & -- & 0 & 6.265s & 2 & (4.7) \\ \hline
5th-order mKdV (Ito) & 0.438s & 4 & 1.000s & 4 & (4.11) \\ \hline
7th-order mKdV (Ito) & 10.469s & 4 & 5.531s & 4 & (4.13) \\ \hline
Generalized Fisher   & 0.406s & 4 & 0.469s & 2 & (5.1) \\ \hline
Nonlinear heat conduction & -- & 0 & 0.485s & 2 & (5.3) \\ \hline
Gen.\/ combined KdV-mKdV & -- & 0 & 2.062s & 2 & (5.5) \\ \hline
Boussinesq & 0.218s & 1 & 0.142s & 1 & (4) \\ \hline
KdV & 0.125s & 1 & 0.126s & 1 & (48) \\ \hline
KdV-Zakharov-Kuznetsov & 0.469s & 1 & 0.142s & 1 & (78) \\ \hline
mKdV-Zakharov-Kuznetsov & 0.282s & 2 & 0.642s & 4 & (81) \\ \hline
3D-mKdV & 0.078s & 2 & 1.874s & 2 & (87) \\ \hline
Gen.\/ Kuramoto-Sivashinsky & 0.734s & 16 & 1.453s & 8 & (89) \\ \hline
Fisher & 0.234s & 8 & 0.343s & 4 & (99) \\ \hline
FitzHugh-Nagumo & 0.719s & 12 & -- & 0 & (101) \\ \hline
Combined KdV-mKdV & 0.204s & 2 & 0.251s & 2 & (109) \\ \hline
Duffing & 0.094s & 4 & -- & 0 & (114) \\ \hline
\end{tabular}
\caption{Comparison between codes PDESpecialSolutions.m and RATH.
Test runs performed on a Dell Dimension 8200 PC, 
with 2.40 GHz Pentium 4 processor, 
512 MB of RAM, with Mathematica v.\ 4.1 and Maple v.\ 7.0.
The first 8 equations appear in Li and Liu (2002);
the last 10 equations are listed in this paper.}
\label{tbl:comparison}
\vspace{-0.65cm}
\end{table}
Liu and Li (2002a) present the {\em Maple} code {\rm AJFM} to 
automate the Jacobi elliptic function method for single PDEs.
This package seeks solutions of the form (\ref{cnsncombination}).

The codes {\rm RATH} and {\rm AJFM} use the Ritt-Wu characteristic sets 
method, implemented by Wang (2001ab). 
The {\rm CharSets} package, available in Maple (Wang, 2002), 
is more versatile and powerful than our algorithm in 
Section~\ref{analyzeandsolve}.

Lastly, Abbott {\em et al.\/} (2002) designed a {\em Mathematica} notebook 
with key functions for the computation of polynomial solutions in 
$\sn$ and $\cn.$

There are several symbolic tools for reducing and solving parameterized 
nonlinear algebraic systems. 
Some are part of codes to simplify overdetermined ODE and PDE systems. 
For example, the {\rm Maple} package \verb|Rif| by Wittkopf and Reid (2003)
allows for the computation of solution branches of nonlinear 
algebraic systems. 
The most powerful algebraic solvers use some flavor of the Gr\"obner basis 
algorithm. 
For up-to-date information on developments in this area we refer to 
Grabmeier {\em et al.\/} (2003).
% 
%%%%%%%%%%%%%%%%%%%%%%%%%%%%%%%%Conclusions%%%%%%%%%%%%%%%%%%%%%%%%%%%%%%
% 
\noindent
\vspace{-0.25cm}
\section{Discussion and Conclusions}
\label{conclusions}

We presented several straightforward algorithms to compute special 
solutions of nonlinear PDEs, without using explicit integration. 
We designed the symbolic package \verb|PDESpecialSolutions.m| 
to find solitary wave solutions of nonlinear PDEs involving 
$\tanh, \sech, \cn$ and $\sn$ functions. 

While the software reproduces the known (and also a few presumably new) 
solutions for many equations, there is no guarantee that the code 
will compute the {\rm complete} solution set of {\rm all} polynomial 
solutions involving the $\tanh$ and/or $\sech$ functions, especially 
when the PDEs have parameters. 
This is due to restrictions on the form of the solutions and the 
limitations of the algebraic solver. 

There is so much freedom in mixed tanh-sech solutions that 
the current code is limited to quadratic solutions.

Furthermore, the nonlinear constraints which arise in solving the
nonlinear algebraic system may be quintic or of higher degree, and 
therefore unsolvable in analytic form. 
Also, since our software package is fully automated, it may not return the 
solutions in the simplest form. 

The example in Section~\ref{examplefifthKdV} illustrates this situation. 
By not solving quadratic or cubic equations explicitly the solutions
(computed interactively with {\em Mathematica}) can be presented in 
a more compact and readable form.

In an attempt to avoid the explicit use of {\it Mathematica}'s 
\verb|Solve| and \verb|Reduce| functions, we considered various alternatives. 
For example, we used (i) variants of Gr\"obner bases on the complete system, 
and (ii) combinatorics on the coefficients in the polynomial solutions
(setting $a_{ij}=0$ or $a_{ij} \ne 0,$ for the admissible $i$ and $j).$
None of these alternatives paid off for systems with parameters.

Often, the nonlinear solver returns constraints on the wave parameters
$c_j$ and the external parameters. 
In principle, one should verify whether or not such constraints affect 
the results of the previous steps in the algorithm. 
In particular, one should verify the consistency with the results 
from step 2 of the algorithms. 
We have not yet implemented this type of sophistication. 
% 
%%%%%%%%%%%%%%%%%%%%%Acknowledgements%%%%%%%%%%%%%%%%%%%%%%%%
% 
\section*{Acknowledgements}
\label{acknowledgements}

This material is based upon work supported by the National Science 
Foundation (NSF) under Grants Nos.\ DMS-9732069, DMS-9912293 and CCR-9901929.
Any opinions, findings, and conclusions or recommendations expressed in this
material are those of the authors and do not necessarily reflect the views
of NSF. 

WH thankfully acknowledges the hospitality and support of the Department of 
Applied Mathematics of the University of Stellenbosch, South Africa, 
during his sabbatical visit in Spring 2001. 
Part of the work was done at Wolfram Research, Inc., while
WH was supported by a Visiting Scholar Grant in Fall 2000. 

M.\ Hickman is thanked for his help with {\em Maple}.
E.\ Parkes and B.\ Duffy are thanked for sharing their code {\rm ATFM}.
P.\ Abbott is thanked for sharing his {\em Mathematica} notebooks and testing 
earlier versions of our code.
Z.-B.\ Li and Y.-P. Liu are thanked for providing us with the 
codes {\rm RATH} and {\rm AJFM}.

The authors are grateful to B.\ Deconinck, B.\ Herbst, P.G.L.\ Leach, 
W.\ Malfliet, J.\ Sanders, and F.\ Verheest for valuable comments.
Last but not least, students 
S.\ Nicodemus, P.\ Blanchard, J.\ Blevins, 
S.\ Formaneck, J.\ Heath, B.\ Kowalski, A.\ Menz, J.\ Milwid, and
M.\ Porter-Peden are thanked for their help with the project.  

% 
%%%%%%%%%%%%%%%%%%%%%%Using the software%%%%%%%%%%%%%%%%%%%%%%%%%%%%%%
% 
\vfill
\newpage
\noindent
\section*{Appendix A: Using the software package}
We illustrate the use of the package \verb|PDESpecialSolutions.m| on a PC. 
Users should have access to {\em Mathematica} v.\ 3.0 or higher. 

Put the package in a directory, say myDirectory, on drive C. 
Start a {\em Mathematica} notebook session and execute the commands:
\vskip 1pt
\noindent
\begin{verbatim}
In[1] = SetDirectory["c:\\myDirectory"];  (* specify directory *)

In[2] = <<PDESpecialSolutions.m           (* read in package   *)

In[3] = PDESpecialSolutions[
        {D[u[x,t],t]-alpha*(6*u[x,t]*D[u[x,t],x]+D[u[x,t],{x,3}])+
         2*beta*v[x,t]*D[v[x,t],x] == 0, 
         D[v[x,t],t]+3*u[x,t]*D[v[x,t],x]+D[v[x,t],{x,3}] == 0}, 
         {u[x,t],v[x,t]}, {x,t}, {alpha, beta}, Form -> Sech, 
         Verbose -> True, InputForm -> False, NumericTest -> True, 
         SymbolicTest -> True, SolveAlgebraicSystem -> True 
         (*, DegreeOfThePolynomial -> {m[1] -> 2, m[2] -> 1} *)]; 
\end{verbatim}
\vskip 1pt
\noindent 
The package will compute the $\sech$ solutions (\ref{ckdvsechsol1}) and
(\ref{ckdvsechsol2}) of the coupled KdV equation (\ref{orgckdv}).

If the \verb|DegreeOfThePolynomial| 
$\rightarrow \{ m[1] \rightarrow 2, m[2] \rightarrow 1 \} $ 
were specified, the code would continue with this case only and compute 
(\ref{ckdvsechsol1}). 

If \verb|SolveAlgebraicSystem| $\rightarrow$ \verb|False|, the algebraic 
system will be generated but not automatically solved.

The format of \verb|PDESpecialSolutions| is similar to the {\em Mathematica}
function DSolve.
The output is a list of sub-lists with solutions and constraints. 
The Backus-Naur Form of the function is:
\begin{eqnarray*}
\langle Main\, Function 
\rangle & \to & \mathtt{PDESpecialSolutions}[\langle Equations \rangle, 
\langle Functions \rangle, \\* 
& & \qquad \langle 
Variables \rangle, \langle Parameters \rangle, 
\langle Options \rangle] \\ 
\langle Options \rangle & \to & 
\verb+Form+\rightarrow \langle Form \rangle \; | \; 
\verb|Verbose|\rightarrow \langle Bool \rangle \; | \\*
&& \verb|InputForm|\rightarrow \langle Bool \rangle \; | \\*
&& \verb|DegreeOfThePolynomial| \rightarrow 
\langle List\, of\, Rules \rangle \; | \\*
&& \verb|SolveAlgebraicSystem|\rightarrow \langle Bool \rangle \; | \\*
&& \verb|NumericTest| \rightarrow \langle Bool \rangle \; 
| \; \verb|SymbolicTest| \rightarrow \langle Bool \rangle \\
\langle Form \rangle & \to & \mathtt{Tanh} 
\; | \; \mathtt{Sech} \; | \;\mathtt{SechTanh} 
\; | \; \mathtt{JacobiCN} \; | \; \mathtt{JacobiSN} \\
\langle Bool \rangle & \to & \mathtt{True} \; | \; \mathtt{False} \\
\langle List\, of\, Rules \rangle & \to & \mathtt{
\{ m[1] \rightarrow Integer, m[2] \rightarrow Integer,... \} } 
\end{eqnarray*}
\vskip .001pt
\noindent
The default value of \verb|Form| is \verb|Tanh|.
The package \verb|PDESpecialSolutions.m| has been tested on both UNIX work 
stations and PCs with {\it Mathematica} versions 3.0, 4.0 and 4.1. 
A test set of over 50 PDEs and half a dozen ODEs was used. 
\label{lastpage}
\end{document}